\documentclass[12pt]{article}

\usepackage{jheppub}
\makeatletter
\def\@fpheader{\relax}
\makeatother

\usepackage{amsthm,amsmath,amssymb,mathtools}
\usepackage{mathrsfs}
\usepackage{bm}
\usepackage{algorithm}
\usepackage{qtree}
\usepackage{bbm}
\usepackage{algpseudocode}
\usepackage{graphicx}
\usepackage{wrapfig}
\usepackage{float}
\usepackage{caption}
\usepackage{subcaption}
\usepackage[most]{tcolorbox}
\usepackage[mathscr]{euscript}
\usepackage{dsfont}
\usepackage[mathscr]{euscript}
\usepackage{mathrsfs}
\usepackage{hyperref}
\usepackage{tikz}
\usepackage{booktabs}
\usepackage[hang,flushmargin]{footmisc}
\newcommand\blfootnote[1]{% 
	\begingroup 
	\renewcommand\thefootnote{}\footnote{#1}% 
	\addtocounter{footnote}{-1}% 
	\endgroup 
}
\newenvironment{nohyphens}{%
	\hyphenpenalty=10000
	\exhyphenpenalty=10000
	\sloppy %
}{\par}

\title{Machine learning modularity}

\author[\star,\diamond,\bullet]{Yi Fan}
\author[\dagger,\ddagger,\star,\diamond]{\!\!, Vishnu Jejjala}
\author[\star,\diamond]{\!\!, Yang Lei\blfootnote{\textsuperscript{*}\, The authors are ordered alphabetically and should all be viewed as co-first authors. }}

\affiliation[\,\star]{School of Physical Science and Technology,   Soochow University, 333 Ganjiang Road, \\ Suzhou 215006, P.R.~China}
\affiliation[\,\diamond]{Institute for Quantum Science, Soochow University, 333 Ganjiang Road, \\ Suzhou 215006, P.R.~China}
\affiliation[\,\bullet]{Center for Astronomy and Astrophysics, Department of Physics, Fudan University, \\ Shanghai 200438, P.R.~China}
\affiliation[\,\dagger]{Mandelstam Institute for Theoretical Physics, School of Physics and NITheCS, University of the Witwatersrand, Johannesburg 2050, South Africa}
\affiliation[\,\ddagger]{NSF AI Institute for Artificial Intelligence and Fundamental Interactions (IAIFI) and\\ Department of Physics, Northeastern University, Boston, MA 02115, USA}

\emailAdd{yfan26@m.fudan.edu.cn}
\emailAdd{v.jejjala@wits.ac.za}
\emailAdd{leiyang@suda.edu.cn}

\abstract{\begin{nohyphens}
Based on a transformer based sequence-to-sequence architecture combined with a dynamic batching algorithm, this work introduces a machine learning framework for automatically simplifying complex expressions involving multiple elliptic Gamma functions, including the $q$-$\theta$ function and the elliptic Gamma function.
The model is trained to apply algebraic identities, particularly the SL$(2,\mathbb{Z})$ and SL$(3,\mathbb{Z})$ modular transformations, to reduce heavily scrambled expressions to their canonical forms. 
Experimental results show that the proposed framework achieves high accuracy on in-distribution tests and maintains robust performance under extrapolation, including substantially deeper scrambling depths than those seen during training.
Our results demonstrate that modern sequence-to-sequence Transformer models can identify canonical representatives of expressions related by modular identities, providing a practical framework for automating symbolic manipulations involving special functions arising in supersymmetric quantum field theory and string theory.
\end{nohyphens}
}

\date{}

\begin{document} 
	
\maketitle
	
\section{Introduction}

Modular properties of partition functions serve as a powerful toolkit for understanding the microscopic states underlying black hole entropy in AdS.
Cardy~\cite{Cardy:1986ie} demonstrated the SL$(2,\mathbb{Z})$ modular invariance of the partition function in two-dimensional CFTs:
\begin{equation}\label{eq:2dmodular-inv}
	Z_0 \left[-\frac{1}{\tau}\right] = Z_0 [\tau] \,, \qquad Z_0[\tau] = \text{Tr}(q^{L_0-\frac{c}{24}}) \,, \qquad q= e^{2\pi i \tau} \,.
\end{equation}
This invariance relates the high-temperature phase (the AdS black hole) in the $\tau\to 0$ limit to the low-temperature phase (thermal AdS) in the $\tau \to i\infty$ limit, leading to the Cardy formula that accounts for the entropy of the BTZ black hole in the dual theory~\cite{Strominger:1996sh,Strominger:1997eq}.

Moreover, the modularity of the partition function on $T^2$ --- exemplified by~\eqref{eq:2dmodular-inv} --- admits a geometric interpretation: conformality implies that the physics depends only on the shape of the torus, parameterized by $\tau$.
Such modular structures have been generalized to other classes of modular forms, including Jacobi forms, mock theta functions, and Igusa cusp forms, which constitute the building blocks of partition functions for two-dimensional superconformal field theories (SCFTs)~\cite{Dabholkar:2012nd}.

Modular structures in higher-dimensional conformal field theories are even richer.
For instance, superconformal indices of four-dimensional $\mathcal{N}=1$ SCFTs have been extensively used to study the microscopic states of BPS black holes in AdS$_5$~\cite{Kinney:2005ej, Hosseini:2017mds, Cabo-Bizet:2018ehj, Choi:2018hmj, Benini:2018ywd, Goldstein:2020yvj}.
In all the investigated in four-dimensional supersymmetric theories, the state counting ultimately follows from the SL$(3,\mathbb{Z})$ modular properties of the elliptic Gamma function~\cite{Ruijsenaars1997,Felder_2000}:
\begin{align} \label{eq:Qfeldermoduality}
	\begin{split}
\Gamma(z;\tau,\sigma) &= e^{-i\pi Q(z,\tau,\sigma)} \Gamma \left(\frac{z}{\tau};-\frac{1}{\tau} , \frac{\sigma}{\tau}\right) \Gamma\left(
\frac{z}{\sigma};-\frac{1}{\sigma} , \frac{\tau}{\sigma}
\right) \,, \\
	Q(z;\tau,\sigma) &= \frac{z^3}{3\tau\sigma} -\frac{\tau+\sigma-1}{2\tau\sigma} z^2 + \frac{\tau^2+\sigma^2 +3\tau \sigma -3\tau-3\sigma+1}{6\tau\sigma} z \\
&+ \frac{(\tau+\sigma-1)(\tau^{-1}+\sigma^{-1}-1)}{12} \,.
	\end{split}
\end{align}
This function coincides with the supersymmetric partition function of a chiral multiplet in $\mathcal{N}=1$ theories.
The modular transformation~\eqref{eq:Qfeldermoduality} can be interpreted as a holomorphic factorization of the supersymmetric partition function in four dimensions~\cite{Nieri:2015yia, Yoshida:2014qwa, Peelaers:2014ima,Gadde:2020bov}, where the two elliptic Gamma functions on the right-hand side correspond to partition functions on $D_2\times T^2$~\cite{Longhi:2019hdh}.
Geometrically, this factorization arises from a Heegaard splitting~\cite{Gadde:2020bov}.
More general SL$(3,\mathbb{Z})$ transformations, defined by arbitrary SL$(3,\mathbb{Z})$ matrices extending~\eqref{eq:Qfeldermoduality}, have been employed to analyze the growth of degeneracies at roots of unity saddles~\cite{Cabo-Bizet:2019eaf, Cabo-Bizet:2020nkr, Jejjala:2021hlt,Aharony:2021zkr,Colombo:2021kbb,Cabo-Bizet:2021jar,Cabo-Bizet:2021plf,Choi:2023tiq,Aharony:2024ntg}. 
Physically, this reflects the ambiguity in choosing the thermal cycle within the Heegaard splitting of a Lens space $L(p,q)\times S^1$~\cite{Jejjala:2022lrm}.

Modular properties of partition functions in non-supersymmetric conformal field theories are less understood.
For free CFTs on $S^{d-1}\times S^1$, modular features were uncovered by studying Mellin transforms of the partition functions, which translate modular invariance into reflection formulas for the Mellin images.
However, on $S^3\times S^1$ the modular invariant quantity is the differentiation of free energy, which transforms as weight $4$ modular form~\cite{cardy1991operator, Kutasov:2000td, Gibbons:2005vp}.
In three-dimensional CFTs the modular-invariant object involves non-local transformations of the thermal partition function~\cite{cardy1991operator, Oshima:1992yy}.
This limitation reveals complexity and fruitfulness of the modular structure of the partition functions.

Recent progress~\cite{Lei:2024oij} has approached a simpler question: given a computed partition function, what is the natural form of its modular properties based on its functional characteristics?
It was found that for CFTs on $S^{d-1}\times S^1$ (with even $d$), the partition function can be expressed as a multiple elliptic Gamma function of rank $d-1$~\cite{Nishizawa_2001}, which exhibits SL$(d+1, \mathbb{Z})$ modular transformation following~\cite{narukawa2004modular}.
Other examples include CFTs on $T^d$, whose partition functions display SL$(d,\mathbb{Z})$ invariance due to the exchange of $S^1$ cycles within the torus~\cite{Shaghoulian:2016gol, Aggarwal:2024axv}.

These examples illustrate diverse modular behaviors, characterized by two key ingredients: the modular group and the degree of the automorphic form.
Even for the same SL$(2,\mathbb{Z})$ group, the partition function of a three-dimensional SCFT can transform via holomorphic factorization, mirroring the Heegaard splitting of the underlying three-manifold~\cite{Nieri:2015yia}.
Conversely, functions with one holomorphic and two elliptic variables may transform differently.
For example, the elliptic Gamma function $\Gamma(z;\tau,\sigma)$ transforms as in~\eqref{eq:Qfeldermoduality} under SL$(3,\mathbb{Z})$, reflecting its nature as an automorphic form of degree $1$.
In contrast, the partition function of a two-dimensional CFTs with holomorphic and anti-holomorphic sectors remains invariant under SL$(2,\mathbb{Z})$ transformations.\footnote{
The $z\to 0$ limit of $\Gamma(z;\tau,\sigma)$ factorizes into holomorphic and anti-holomorphic parts upon identifying $\sigma =\bar{\tau}$~\cite{Lei:2024oij}.}
Such rich modular identities are essential for identifying the nature of the underlying automorphic forms, yet they are often difficult to uncover.
Equivalently, one may ask: given a generic function, how can one determine its relevant modular group and transformation law?

Machine learning has recently attracted attention in string theory as a powerful tool for predicting unknown physics from large datasets~\cite{He:2017set,Krefl:2017yox,Ruehle:2017mzq,Carifio:2017bov}.
This idea has subsequently been applied to holographic settings.
For representative papers, see~\cite{Hashimoto:2018ftp,Hashimoto:2018bnb,Hu:2019nea,Hashimoto:2019bih,Akutagawa:2020yeo,Song:2020agw,Park:2022fqy,Park:2023slm,Jejjala:2023zxw,Bea:2024xgv,Jejjala:2025hgv,Kim:2025aln,Jeong:2025omu}.
In our context, we wish to explore whether machine learning can infer modular-like properties of functions and discover possible modular identities.
A prerequisite is to test whether machine learning can learn known SL$(r,\mathbb{Z})$-invariant modular functions, as well as simple modular-covariant examples, such as the (multiple)-elliptic Gamma function.
The work~\cite{Dersy:2022bym} demonstrated that machine learning can learn the identities of dilogarithm and trilogarithm functions and simplify expressions involving them with high accuracy~\cite{DBLP:journals/corr/abs-1912-01412}.
This machine learning technique also simplifies expressions including integration by parts~\cite{vonHippel:2025okr} or solves differential equations~\cite{DBLP:journals/corr/abs-1912-01412}.
The transformer architecture has also been applied to study scattering amplitudes in four-dimensional $\mathcal{N}=4$ super-Yang--Mills theory~\cite{Cheung:2024svk,Cai:2024znx,Cai:2025atc}.
Since $q$-$\theta$ functions are built from $q$-Pochhammer symbols --- the $q$-analogues of polylogarithms --- it is natural to ask whether machines can learn general modular identities as a generalization of the work~\cite{Dersy:2022bym}.
This could have potential applications for understanding Farey-tail-like configurations in AdS/CFT~\cite{Dijkgraaf:2000fq, Manschot:2007ha}.

In this work, we aim to address the aforementioned foundational questions through concrete examples. Our investigation by machine learning consists of two complementary parts: the study of modular transformations acting on the relevant moduli parameters and the simplification of expressions involving multiple elliptic Gamma functions under these modular transformations.
By doing so, we demonstrate that machine learning can accurately identify M\"obius transformations. 
Our approach tests whether generic points in the upper half-plane can be mapped into the fundamental domain, with the corresponding SL$(2,\mathbb{Z})$ matrices determined algorithmically~\cite{random_integer_matrix2013}.
Also, we examine how expressions containing multiple elliptic Gamma functions simplify under general actions including duplications and SL$(r,\mathbb{Z})$ transformations.

This paper is organized as follows. 
In Section~\ref{sec:function}, we will discuss the properties of the polylogarithm functions and generalized $q$-Pochhammer symbol, together with the multiple elliptic Gamma functions built on it. 
In Section~\ref{sec:ML-SL2Z}, we will perform numerical analysis on the modulus being acted by the SL$(2,\mathbb{Z})$ transformation.
In Section~\ref{sec:ML-modularfunction}, we explain our algorithm to use machine learning to simplify expressions involving $q$-$\theta$ and elliptic Gamma functions. 
High accuracy is obtained in the numerical tests presented below.

\section{Polylogarithm and generalized $q$-Pochhammer symbol}
\label{sec:function}

\subsection{Polylogarithm}

The polylogarithm can be defined using series as 
\begin{equation}
	\text{Li}_r(x) = \sum_{n=1}^\infty \frac{x^n}{n^r} \,, \qquad |x|<1 \,, \qquad r=1,2, \cdots \,.
\end{equation}
Also when $r=1$, this function reduces to Li$_1(x) = - \ln(1-x)$.
This function can be analytically continued to the whole complex plane $\mathbb{C}$ with the branch cut on $[1,\infty)$. 
This class of functions can also be defined recursively as 
\begin{equation}
	\text{Li}_r (x) = \int_0^x dt\ \frac{\text{Li}_{r-1} (t) }{t} \,.
\end{equation}
We are especially interested in Li$_2(x)$ which satisfies functional identities including~\cite{zagier2007dilogarithm,kirillov1995dilogarithm, Dersy:2022bym}:
\begin{align}\label{eq:dilog-identities}
	\begin{split}
\text{Reflection}:& \quad \text{Li}_2(1-x) = -\text{Li}_2(x) + \frac{\pi^2}{6} - \ln x \ln(1-x) \,, \\
\text{Inversion}:& \quad \text{Li}_2\left(\frac{1}{x}\right) = -\text{Li}_2(x) -\frac{\pi^2}{6} - \frac{1}{2}\ln^2(-x) \,, \\
\text{Duplication}:& \quad \frac{1}{2}\text{Li}_2(x^2) =  \text{Li}_2(x)+\text{Li}_2(-x) \,.
	\end{split}
\end{align}
Such identities originate from considerations of mathematical problems, such as XXZ model~\cite{kirillov1985exact}.
They are crucial in simplifying scattering amplitudes computed from quantum field theories, revealing singularity structures of propagators and correlation functions. 

The pentagon identity of the dilogarithm function is also known as the master identity to generate all three identities in~\eqref{eq:dilog-identities}~\cite{zagier2007dilogarithm}: 
\begin{align}\label{eq:pentagon-identityLi2}
	\begin{split}
& \text{Li}_2(x) +\text{Li}_2(y) + \text{Li}_2 \left(\frac{1-x}{1-xy}\right)
+\text{Li}_2(1-x y)  + \text{Li}_2 \left(\frac{1-y}{1-xy}\right) \\
=& \frac{\pi^2}{6} - \ln x \ln(1-x) - \ln y \ln(1-y)  + \ln  \left(\frac{1-x}{1-xy}\right) \ln \left(\frac{1-y}{1-xy}\right) \,.
	\end{split}
\end{align}
Although there is no known proof that~\eqref{eq:pentagon-identityLi2} is sufficient to derive all possible identities, Goncharov's conjecture states that any dilogarithm identities can be written as linear combinations with rational coefficients of this pentagon identity~\cite{Goncharov1995}.  

\subsection{Multiple $q$-Pochhammer symbols}

We introduce the following notation~\cite{Nishizawa_2001} to parametrize the multiple elliptic Gamma function and the multiple $q$-Pochhammer symbols. Let the fugacities associated with chemical potentials be given by $x = e^{2\pi i z}$ and $q_j = e^{2\pi i \tau_j}$, where $z \in \mathbb{C}$ and $\tau_j \in \mathbb{C} \setminus \mathbb{R}$. 
Define (assume $r\ge 0$)
\begin{align}\label{eq:qs-notation}
	\begin{split}
\underline{q} &:= (q_0,\cdots,q_r) \,, \\
\underline{q}^-(j) &:= (q_0,\cdots,\check{q}_j,\cdots, q_r) \,, \\
\underline{q}[j] &:= (q_0,\cdots,q_j^{-1},\cdots, q_r) \,, \\
\underline{q}^{-1} &:= (q_0^{-1},\cdots, q_r^{-1}) \,,
	\end{split}
\end{align}
where $\check{q}_j$ denotes omission of the $j$-th component.
The same convention~\eqref{eq:qs-notation} applies to the chemical potentials $\tau_j$:
\begin{align}\label{eq:taus-notation}
	\begin{split}
\underline{\tau} &:= (\tau_0,\cdots,\tau_r) \,, \\
\underline{\tau}^-(j) &:= (\tau_0,\cdots,\check{\tau}_j,\cdots, \tau_r) \,, \\
\underline{\tau}[j] &:= (\tau_0,\cdots,-\tau_j,\cdots, \tau_r) \,, \\
-\underline{\tau} &:=(-\tau_0,\cdots,-\tau_r) \,.
	\end{split}
\end{align}

For $\mathrm{Im}(\tau_j) > 0$, the multiple $q$-Pochhammer symbol is defined as
\begin{equation}
	(x; \underline{q} )_{\infty}^{(r)} := \prod_{j_0,\cdots,j_r=0}^\infty (1-x q_0^{j_0} \cdots q_r^{j_r}) \,.
\end{equation}
This definition extends to regimes where $\mathrm{Im}(\tau_j)<0$ for $j=0,\dots,k-1$ and $\mathrm{Im}(\tau_j)>0$ for $j=k,\dots,r$ via the prescription
\begin{equation}
	(x;\underline{q})_{\infty}^{(r)} := \left[ 
\prod_{j_0,...,j_r=0}^\infty (1-x q_0^{-j_0-1}	\cdots q_{k-1}^{-j_{k-1}-1} q_k^{j_k}\cdots q_r^{j_r} )
	\right]^{(-1)^k} \,.
\end{equation}
The generalized $q$-Pochhammer symbol is then used to define the multiple elliptic Gamma functions:
\begin{equation}\label{eq:def-multipleelliptic}
	G_r(z|\underline{\tau}) := (x^{-1}q_0\cdots q_r;\underline{q})_\infty^{(r)} \left[(x;\underline{q})_{\infty}^{(r)} \right]^{(-1)^r} \,.
\end{equation}
The well-known $q$-$\theta$ function $\theta(z;\tau)$ and the elliptic Gamma function $\Gamma(z;\tau,\sigma)$~\cite{Ruijsenaars1997,Felder_2000,Felder_2008} correspond respectively to the cases $r=0$ and $r=1$ of $G_r(z|\underline{\tau})$, \textit{i.e.},
\begin{align}
	\begin{split}
&	G_0(z|\tau)= \theta(z;\tau) \,, \qquad G_1(z|\tau,\sigma) = \Gamma(z;\tau,\sigma) \,.
	\end{split}
\end{align}
The $\theta(z;\tau)$ appears in many physical models including the partition functions on $T^2\times S^2$~\cite{Closset:2013sxa, Honda:2015yha} or also partition functions on two dimensional supersymmetric field theories. 
The $\Gamma(z;\tau,\sigma)$ are partition functions of $\mathcal{N}=1$ chiral multiplet in $S^3\times S^1$~\cite{Peelaers:2014ima}. 
And higher ranks of elliptic Gamma appear in chiral multiplet in 6d SCFT or 5d SYM theory~\cite{Qiu:2015rwp}. 
For applications of these functions, see~\cite{Tizzano:2014roa, Winding:2016wpw}. 

The multiple elliptic Gamma functions defined in~\eqref{eq:def-multipleelliptic} possess several remarkable properties~\cite{narukawa2004modular}:
\begin{itemize}
	\item \textbf{Shifts}: There are $(r+1)$ different shifts in $\tau_j$.
	\begin{align} \label{eq:shiftsGamma}
		\begin{split}
G_r(z+1|\underline{\tau}) &=G_r(z|\underline{\tau}) \,,\\
G_r(z+\tau_j|\underline{\tau}) &=G_r(z|\underline{\tau}) G_{r-1}(z|\underline{\tau}^-(j)) \,, \qquad r\ge 1\,, \\
G_r(z|\underline{\tau}) &= \frac{1}{G_r(z-\tau_j| \underline{\tau}[j])} \,.
\end{split}
	\end{align}
\item \textbf{Inversion}: The transformation $z \to -z$ (equivalently $x \to x^{-1}$) yields:
\begin{equation} \label{eq:inversion}
G_r(-z|-\underline{\tau}) = \frac{1}{G_r(z|\underline{\tau})} \,.
\end{equation}
\item \textbf{SL$(r,\mathbb{Z})$ modularity}: The modular properties can be expressed in two ways. First, a relation among the functions themselves:
\begin{equation}\label{eq:Modularity-Gammar}
	\prod_{k=1}^r G_{r-2} \left(\frac{z}{\omega_k}\Big| \frac{\omega_1}{\omega_k},\cdots, \frac{\check{\omega}_k}{\omega_k},\cdots \frac{\omega_r}{\omega_k} \right) = \exp\left[ -\frac{2\pi i}{r!} B_{rr} (z|\underline{\omega})\right] \,,
\end{equation}
where $B_{r,n}(z|\underline{\omega})$ are Bernoulli polynomials defined via the generating function
\begin{equation}\label{eq:def-Bernoulli-poly}
\frac{t^r e^{zt}}{\prod_{j=1}^r (e^{\omega_j t}-1)} = \sum_{n=0}^\infty B_{r;n}(z|\underline{\omega}) \frac{t^n}{n!} \,.
\end{equation}
A second formulation involves the multiple sine function $S_r(z|\underline{\omega})$~\cite{kurokawa2003multiple}:
\begin{align}\label{eq:product-Gamma-Sine}
	\begin{split}
G_r(z|\underline{\tau}) &= \exp\left[ -
\frac{2\pi i}{(r+2)! } B_{r+2,r+2} (z|\underline{\tau},1)
\right] \\ &\times \prod_{k=0}^\infty \frac{S_{r+1}(z+k+1|\underline{\tau})^{(-1)^r} S_{r+1}(z-k|\underline{\tau})^{(-1)^r}}{\exp \{ \frac{i\pi}{(r+1)!} [ B_{r+1,r+1} (z+k+1|\underline{\tau}) - B_{r+1,r+1} (z-k|\underline{\tau})	]
			\}} \,.
	\end{split}
\end{align}
Here the multiple Sine function $S_r(z| \underline{\omega})$ can be conveniently represented as \cite{narukawa2004modular}: 
\begin{align}
	\begin{split}
S_r(z| \underline{\omega}) &= \exp \left[ (-1)^r \frac{\pi i}{r!} B_{rr}(z|\underline{\omega})\right] \prod_{k=1}^r (x_k ; \underline{q_k})_\infty^{(r-2)} \\
&= \exp \left[ (-1)^{r-1} \frac{\pi i}{r!} B_{rr}(z|\underline{\omega})\right] \prod_{k=1}^r (x_k^{-1} ; \underline{q_k^{-1} })_\infty^{(r-2)} \,,
	\end{split}
\end{align}
where the parameters are defined as
\begin{equation}
x_k = e^{2\pi i \frac{z}{\omega_k}}, \qquad q_k = (e^{2\pi i \frac{\omega_1}{\omega_k}}, \cdots, e^{2\pi i \frac{\omega_{k-1}}{\omega_k}}, e^{2\pi i \frac{\omega_{k+1}}{\omega_k}}, \cdots, e^{2\pi i \frac{\omega_r}{\omega_k}}) \,.
\end{equation}
The function is related to the conventional trigonometric function by $S_1(z|\omega) = 2\sin \frac{\pi z}{\omega}$. 
This formula is used in~\cite{Lei:2024oij} to study modularity of free conformal field theories. 
\item \textbf{Multiplication}~\cite{Lei:2024oij,felder2002multiplication}:
\begin{align}
\begin{split}
	G_r(z|\underline{\tau} ) =\prod_{ \underline{a}=0 }^{m-1} G_r(z+ \underline{a} \cdot \underline{\tau}| m\underline{\tau}+\underline{n}) \,,
\end{split}
\end{align}
where $\underline{a} \cdot \underline{\tau} = \sum_{i=0}^{r} a_i \tau_i$ and $m\underline{\tau}+\underline{n} = (m\tau_0+n_0,\dots,m\tau_r+n_r)$.
\item \textbf{Duplication}: Duplication formulas are also established for the classical cases $\theta(z;\tau)$ and $\Gamma(z;\tau,\sigma)$~\cite{felder2002multiplication}:\footnote{These can be viewed as special instances of the more general \emph{first multiplication formula} presented in~\cite{felder2002multiplication}, which falls outside the scope of this paper.}
\begin{align}\label{eq:duplication}
	\begin{split}
		& \theta(2z;\tau) = \theta(z;\tau)\theta\left(z+\frac{1}{2};\tau\right)
		\theta\left(z+\frac{\tau}{2};\tau\right) \theta\left(z+\frac{\tau+1}{2};\tau\right) \,, \\ 
		& \Gamma(2z;\tau,\sigma) = \Gamma(z;\tau,\sigma) \Gamma\left(z+\frac{1}{2};\tau,\sigma \right) \Gamma\left(z+\frac{\tau}{2};\tau,\sigma \right)
		\Gamma\left(z+\frac{\sigma}{2};\tau,\sigma \right) \\
		& \times \Gamma\left(z+\frac{1+\tau}{2};\tau,\sigma \right) \Gamma\left(z+\frac{\tau+\sigma}{2};\tau,\sigma \right)
		\Gamma\left(z+\frac{\sigma+1}{2};\tau,\sigma \right) \\
		& \times \Gamma\left(z+\frac{\sigma+1+\tau}{2};\tau,\sigma \right) \,.
	\end{split}
\end{align}
For higher-rank multiple elliptic Gamma functions (\textit{i.e.}, $r > 1$), the corresponding duplication formulas become substantially more complex.
\end{itemize}

Formulas exist that combine the inversion symmetry with the $S$-transformation of the $\mathrm{SL}(r,\mathbb{Z})$ action described in~\eqref{eq:Modularity-Gammar} into a unified transformation rule under general matrix elements, at least for the cases $r=0$ and $r=1$.
For the $q$-$\theta$ function $\theta(z;\tau)$, one has
\begin{equation}\label{eq:thetaSL2Z}
	\theta\left(\frac{z}{ m\tau+n};\frac{k\tau+l}{m\tau+n} \right)  =e^{i\pi B_2^\mathbf{m}(z;\tau)} \theta(z;\tau)  \,, \quad \mathbf{m}=(m,n)\,,
\end{equation}
where the phase is expressed in terms of a deformed second Bernoulli polynomial~\cite{Felder_2008,Jejjala:2022lrm} together with $\sigma_k(\vec{n};m)$ denotes the generalized Fourier--Dedekind sum: 
\begin{equation}\label{eq:theB-polynomial}
B_2^\mathbf{m}(z;\tau) = \frac{1}{m} B_{22} \left(m z +1;m\tau+n\right) +2\sigma_1(n,1;m)\,,
\end{equation} 
where the Fourier-Dedekind sum is a generalization of Dedekind sum \cite{rademacher1972dedekind} and is computed by \cite{beck2002frobenius}:  
\begin{equation}\label{eq:definition-sigmat}
	\sigma_t (n_1,n_2 \cdots,n_r;m) = \frac{1}{m} \sum_{\xi^m=1 \neq \xi} \frac{\xi^t}{(\xi^{n_1}-1) \cdots (\xi^{n_r} -1)} \,.
\end{equation}
Together with the shift symmetry, the full modular action on $\tau$ generates the group $\mathrm{SL}(2,\mathbb{Z}) \ltimes \mathbb{Z}^2$.
Similarly, the elliptic Gamma function $\Gamma(z;\tau,\sigma)$ satisfies an $\mathrm{SL}(3,\mathbb{Z})$ modular identity:
\begin{align}\label{eq:SL3ZGamma}
	\begin{split}
		\Gamma(z;\tau,\sigma)&=e^{-i\pi Q_{\mathbf{m}}(z;\tau,\sigma)}\Gamma\left(\tfrac{z}{m\sigma+n};\tfrac{\tau-\tilde{n}(k\sigma+l)}{m\sigma+n},\tfrac{k\sigma+l}{m\sigma+n}\right)\Gamma\left(\tfrac{z}{m\tau+\tilde{n}};\tfrac{\sigma-n(\tilde{k}\tau+\tilde{l})}{m\tau+\tilde{n}},\tfrac{\tilde{k}\tau+\tilde{l}}{m\tau+\tilde{n}}\right)\,,
	\end{split}
\end{align}
where the phase $Q_{\textbf{m}}(z;\tau,\sigma)$ is given by~\cite{Jejjala:2022lrm}
\begin{equation}\label{eq:phase-FHRZ-sumabpart}
	Q_{\textbf{m}}(z;\tau,\sigma)=	\tfrac{1}{m}B_{33}(mz+1;m\tau+\tilde{n},m\sigma+n,1)+2\sigma_1(n,\tilde{n},1;m)\,.
\end{equation}
Combined with the shift symmetry, the complete transformation group is $\mathrm{SL}(3,\mathbb{Z}) \ltimes \mathbb{Z}^3$~\cite{Gadde:2020bov}. 
These modular formulas are particularly useful for extracting the $\mathrm{SL}(3,\mathbb{Z})$ saddle points in the dual gravitational description~\cite{Cabo-Bizet:2019eaf,Jejjala:2021hlt, Jejjala:2022lrm,Lei:2024oij}.

The multiple generalized $q$-polylogarithms considered here are closely related to the generalized $q$-Pochhammer symbol.
We emphasize that the term “multiple” here has a different meaning from that in the multiple elliptic polylogarithms in the literature \cite{Brown:2011bs}.
In the present Barnes-type construction, multiplicity refers to several independent lattice parameters $\tau_j$ (equivalently $q_j$) entering the multiple $q_j$-product, analogous to the multiple lattice periods of the Barnes zeta function 
\begin{equation}
	\zeta_r (s,z|\underline{\tau}) = \sum_{n_i\ge 0} \frac{1}{(z+\sum_{i} n_i \tau_i)^s} \,.
	\end{equation} 
In the Brown–Levin construction \cite{Brown:2011bs}, by contrast, multiplicity refers to polylogarithmic depth, analogous to the indices $(s_1,\ldots,s_k)$ of multiple zeta values, rather than to several independent elliptic parameters. 
Thus, the two constructions implement distinct notions of “multiple” and should not be identified.
 
On the one hand, Nishizawa~\cite{Nishizawa_2001} introduced the multiple generalized $q$-polylogarithm
\begin{equation}
	\text{Li}_{r+2} (x;\underline{q}) = \sum_{n=1}^\infty \frac{x^n}{n \prod_{j=0}^r (1-q_j^n)} \,.
\end{equation}
This function is connected to the $q$-Pochhammer symbol via the relation~\cite{narukawa2004modular}
\begin{equation}
	\text{Li}_{r+2}(x;\underline{q}) =- \ln (x;\underline{q})_{\infty}^{(r)} \,.
\end{equation}
It is therefore natural to ask how identities for polylogarithm functions correspond to those for the generalized $q$-Pochhammer symbol. In the limit $q\to 1$ (\textit{i.e.}, $\tau\to0$), for instance, the inversion and duplication identities reduce precisely to known identities for polylogarithms.
Connections also exist between polylogarithms and elliptic Gamma functions~\cite{Pa_ol_2018}. The unrefined elliptic Gamma function $\Gamma(z;\tau,\tau)$, in particular, can be linked to the dilogarithm. 
The function defined as
\begin{equation}\label{eq:TGamma}
	T(z;\tau) = \tau \ln \Gamma(z+\tau;\tau,\tau) - \ln \Gamma\left(\frac{z-1}{\tau};-\frac{1}{\tau},-\frac{1}{\tau}\right) \,,
\end{equation}
admits the representation 
\begin{align}\label{eq:T=thetaLi2}
	\begin{split}
T(z;\tau)& = \frac{\pi i(\tau-2z)(1+2\tau z-2z^2)}{12 \tau} + z\ln \theta(z;\tau) \\
&- \frac{1}{2\pi i} \sum_{m=0}^\infty \left[ \text{Li}_2 \left(\frac{q^{m+1}}{x}\right) - \text{Li}_2 \left(x q^m\right)\right]\,.
	\end{split}
\end{align}
These are crucial to formulate the elliptic extension of the unrefined elliptic Gamma function, revealing the growth of degeneracy of $\mathcal{N}=4$ SYM near the rational saddles. 

The pentagon identity, serving as the master identity that generates relations for the dilogarithm function, admits a $q$-deformation~\cite{Faddeev:1993rs}, albeit one that is restricted to variables satisfying Weyl relations.
Consequently, it remains unclear how to extend the reflection identity of the dilogarithm Li$_2$ to the  $q$-Pochhammer symbol $(x;q)_\infty$. 
The modular transformations in~\eqref{eq:thetaSL2Z} and~\eqref{eq:SL3ZGamma} are fundamentally tied to operations on the elliptic parameters, which fall outside the scope of identities derivable from the pentagon master identity~\eqref{eq:pentagon-identityLi2}. 
Expressions involving elliptic Gamma functions and higher-rank generalizations introduce new layers of complexity. 
In the next two sections we will explore how to use machine learning to study these modular transformations. 

\section{Machine learning M\"obius transformation}
\label{sec:ML-SL2Z}

The most elementary modular transformation is that of the SL$(2,\mathbb{Z})$ group, which acts on the modulus $\tau$ via
\begin{equation}
	g\cdot \tau = \frac{k \tau+l}{m\tau+n} \,, \qquad kn-ml=1 \,, \quad k,n,m,l\in \mathbb{Z} \,.
\end{equation}
This action preserves the condition that $\tau$ lies in the upper half-plane $\mathbb{H}$.
The standard fundamental domain $\mathcal{F} \subset \mathbb{H}$ for the $\mathrm{SL}(2,\mathbb{Z})$ action is
\begin{equation}
\mathcal{F} = \left\{ z \in \mathbb{H} \,\middle|\, |z| \geq 1,\ -\tfrac{1}{2} \leq \mathrm{Re}(z) \leq \tfrac{1}{2} \right\} \,,
\end{equation}
Any SL$(2,\mathbb{Z})$ matrix can be decomposed into a sequence of the generators $T$ and $S$ by following the Euclidean algorithm. 
Under successive $T$ and $S$ transformations, the fundamental domain is mapped to various copies in $\mathbb{H}$, which together tessellate the entire upper half-plane.

To employ machine learning in studying modular forms and related functions, a preliminary question must be addressed: can machine learning effectively recognize modular transformations?
Given a point in $\mathbb{H}$, an SL$(2,\mathbb{Z})$ transformation can map it to another point lying in some image of the fundamental domain. 
The matrix that connects these two points can be determined algorithmically, for instance via the method described in~\cite{random_integer_matrix2013}.
However, an ambiguity arises in how one chooses the initial point, an issue tied to the measure used for sampling points on $\mathbb{H}$.

The upper half-plane can be mapped conformally to the Poincaré disk by the holomorphic transformation
\begin{equation}\label{eq:coordinate=disktouper}
	z_\mathbb{H} = i\,\frac{1 + w}{1 - w} \,, \qquad w = \tanh \left(\frac{r}{2} \right) e^{i\theta}=u+i v \,,
\end{equation}
where $w$ stands for the coordinates on disk. 
Thus, an equivalent question to sample a point on the upper half plane is: what measure should be assigned to the unit disk? 
The most natural choice on the Poincar\'e disk is the hyperbolic measure, which corresponds to the hyperbolic structure inherent to the moduli space under study. 
Alternatively, one may regard the disk as a conventional Euclidean disk and consider other possible measures. 
This leads to classical constructions such as those appearing in Bertrand's paradox, which proposes various inequivalent notions for selecting a random chord, such as choosing two random points on the circumference and drawing the chord between them, or choosing the midpoint of the chord by a uniform area measure, or choosing the perpendicular to a random point on a random radius~\cite{aerts2014solving}.
In this section, we will examine these four different sampling measures and test how each affects the predictive performance of the corresponding machine‑learning models.

\subsection{Sampling under the various measures}

We will display four different  possible measures on the disk and map it back to the upper half plane. 
Given a density function on the disk $f_{\mathbb{D}}(u,v)$, the density on the upper half plane can be determined by the transformation~\eqref{eq:coordinate=disktouper} as 
\begin{equation}\label{eq:transformation-Jacobian}
f_{\mathbb{H}} (x,y) = 	f_{\mathbb{D}} \Big(u(x,y),v(x,y) \Big) \left| \det \frac{\partial(u,v)}{\partial(x,y)} \right| \,.
\end{equation}

\subsection*{Hyperbolic measure}

\begin{figure}[H]
	\includegraphics[width=\textwidth]{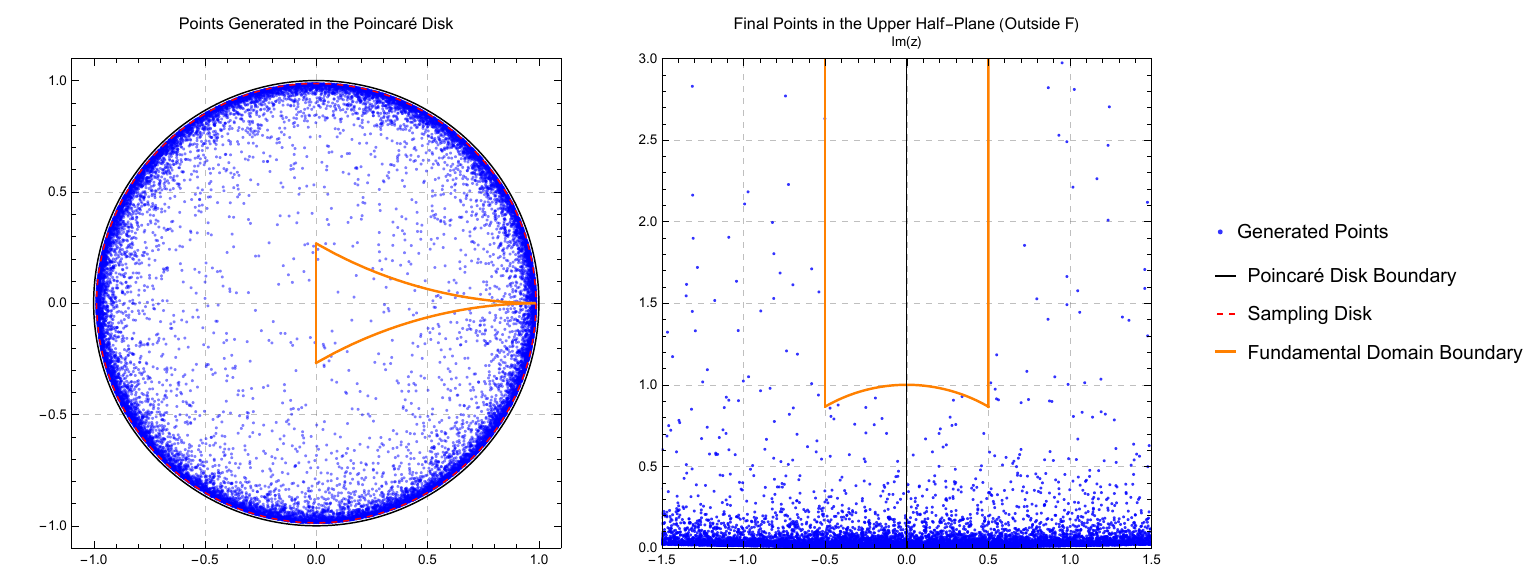}
	\caption{Hyperbolic cutoff radius $R_{h} = 5$, Euclidean radius $r = 0.99$, $N = 20{,}000$.}\label{fig:R5}
\end{figure}

The hyperbolic distance from any interior point of the Poincar\'e disk to its boundary is infinite. To construct a finite data set, we introduce a cutoff radius $R_h$ and generate points only within the region $r \le R_h$. 
These points are then mapped to the upper half-plane using the transformation~\eqref{eq:coordinate=disktouper}. From the resulting set, we retain only those points that lie outside the standard fundamental domain; points inside the domain are excluded from the training data.
This sampling procedure follows the truncated hyperbolic measure. As a consequence, the distribution exhibits a higher density of points near the boundary of the Poincar\'e disk and, correspondingly, near the real axis in the upper half-plane. 
An example of such sample is illustrated in Figure~\ref{fig:R5}.

\paragraph{Bertrand~I: Random chord endpoints.}
Bertrand's original problem concerns the definition of a ``random chord'' in a circle. Here we adapt it to sample points inside the disk by taking the midpoint of the chord selected according to Bertrand's first construction.\footnote{
Approximate Ricci-flat Calabi--Yau metrics have been constructed using physics informed neural networks~\cite{Ashmore:2019wzb,Anderson:2020hux,Douglas:2020hpv,Jejjala:2020wcc}.
The sampling of points on the Calabi--Yau manifolds, following~\cite{Shiffman:1998drw}, is morally akin to Bertrand~I.
It has been hypothesized that a superior point selection scheme would yield faster numerical convergence, and alternatives have been proposed~\cite{BBHJMMT,Ruehle25}.
In this work, we similarly notice sensitivity to the sampling algorithm in the performance.
}
\begin{figure}[H]
	\centering
	\includegraphics[width=0.3\textwidth]{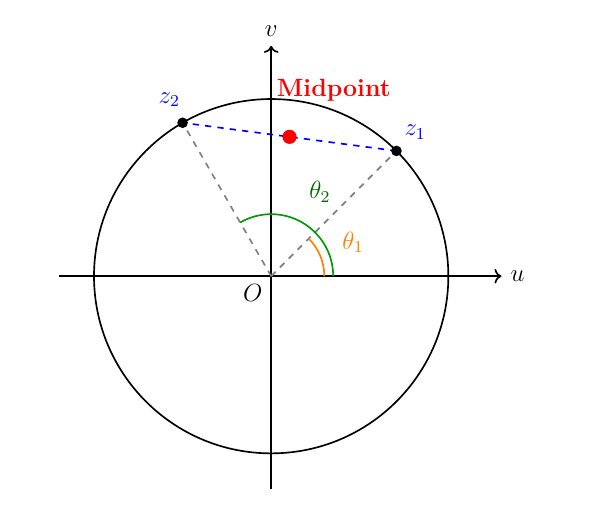}
	\caption{Bertrand~I sampling.}\label{fig:chordal}
\end{figure}
Specifically, the first Bertrand construction proceeds as follows: fix one endpoint of the chord uniformly on the circle, then choose the second endpoint independently and uniformly on the circle. The midpoint of the resulting chord is taken as the sampled point inside the disk, as illustrated in Figure~\ref{fig:chordal}.

\begin{figure}[H]
	\includegraphics[width=\textwidth]{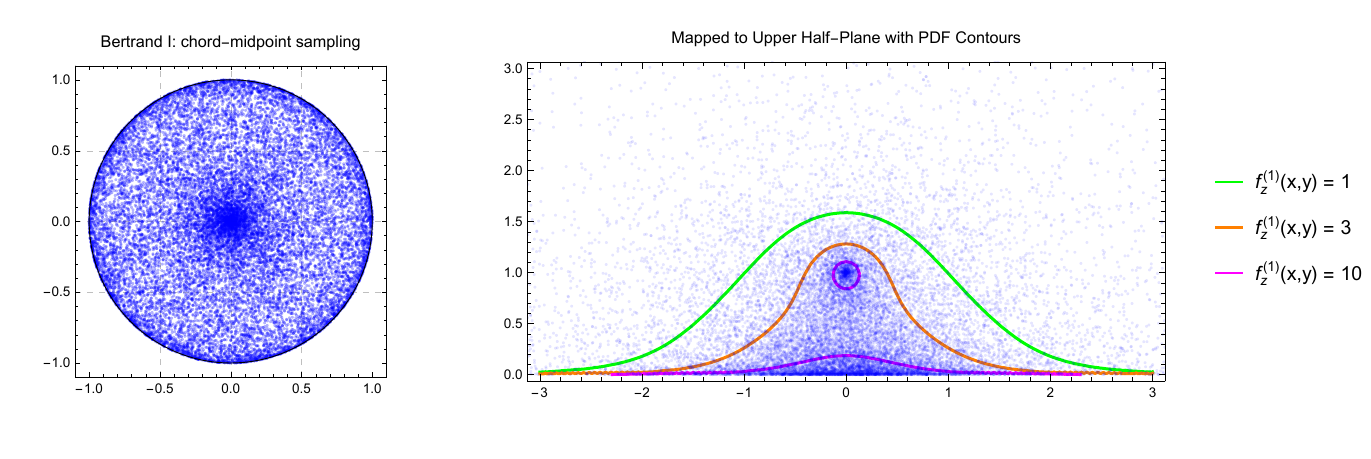}
	\caption{Bertrand~I (chord-midpoint) sampling, $N = 20{,}000$.}\label{fig:chordal_dist}
\end{figure}

This procedure induces an isotropic but non-uniform density in the disk. 
The corresponding probability density function  on $\mathbb{D}$ is~\cite{jaynesWellposedProblem1973}
\begin{equation}
	\,f_{\mathbb{D}}^{\mathrm{I}}(u,v)=\frac{1}{\pi^2\,|w|\,\sqrt{1-|w|^2}} \,.
\end{equation}
The transformation~\eqref{eq:transformation-Jacobian} yields the probability density on the upper half plane $\mathbb{H}$ as
\begin{equation}
	\,f_{\mathbb{H}}^{\mathrm{I}}(x,y)=\frac{2}{\pi^2}\,
	\frac{1}{\sqrt{y}\,\sqrt{x^2+(y-1)^2}\,\big(x^2+(y+1)^2\big)} \,.
\end{equation}
The mapping preserves the qualitative features of the chordal density: points cluster near the real axis and around $z=i$, the image of the disk center. As in the hyperbolic case, points that land inside the fundamental domain are discarded. This distribution is visualized in Figure~\ref{fig:chordal_dist}.

\begin{figure}[H]
	\includegraphics[width=\textwidth]{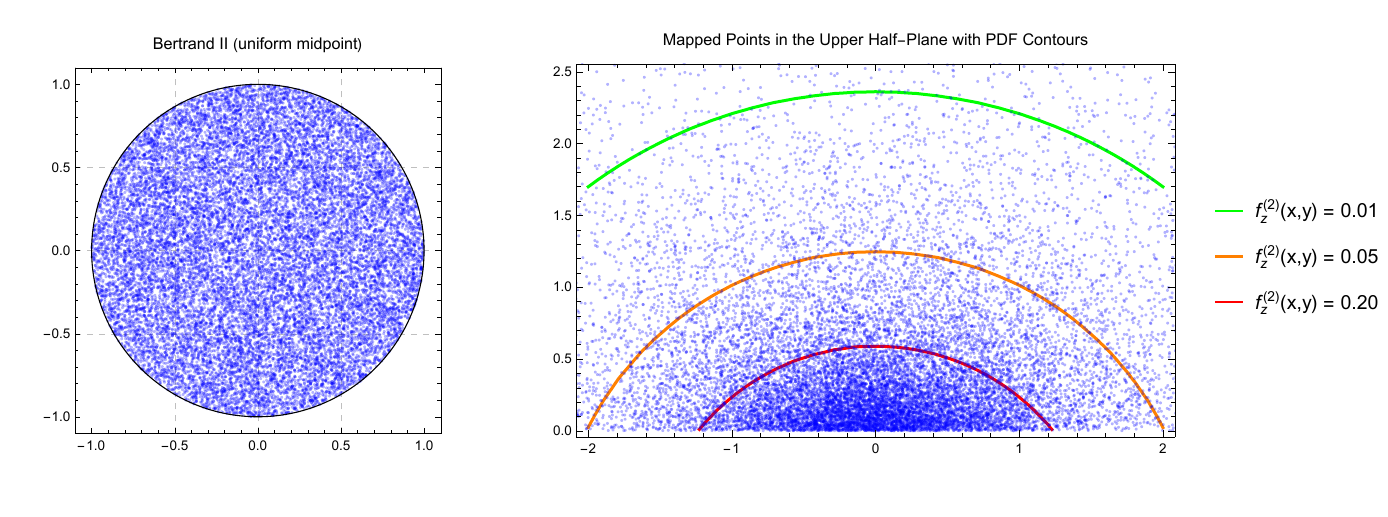}
	\caption{Bertrand~II (Euclidean area-uniform) sampling, $N = 20{,}000$.}\label{fig:euclidean}
\end{figure}

\paragraph{Bertrand~II: Euclidean (area-uniform) sampling.}
Since the Poincar\'e disk has finite Euclidean area, we may sample points uniformly with respect to the standard Euclidean area measure on the unit disk. The corresponding probability density on the disk is simply constant:
\begin{equation}
	\,f_{\mathbb{D}}^{\text{II}}(u,v)=\frac{1}{\pi} \,,\qquad u^2+v^2<1\,.
\end{equation}
However, the M\"obius transformation that maps the disk to the upper half-plane is not an isometry of the Euclidean metric; consequently, the induced distribution on $\mathbb{H}$ becomes non-uniform. 
Using the Jacobian of the conformal map, the probability density function on  $\mathbb{H}$ is obtained as
\begin{equation}
	\,f_{\mathbb{H}}^{\text{II}}(x,y)=\frac{4}{\pi}\,\frac{1}{\big(x^2+(y+1)^2\big)^2} \,.
\end{equation}
This expression clearly shows a strong accumulation of density near the real axis $y=0$ and the power law decay as imaginary parts being large $y\to\infty$. As before, points that fall inside the standard fundamental domain are discarded. 
The resulting distribution is displayed in Figure~\ref{fig:euclidean}.

\paragraph{Bertrand~III: Random radial distance of the chord midpoint.}
The third classical prescription samples a chord by first choosing a direction $\theta\sim\mathrm{Unif}(0,2\pi)$ and then choosing a distance $r\sim\mathrm{Unif}(0,1)$ along the corresponding radius. 
The point $w=r e^{i\theta}$ is taken to be the chord midpoint (with the chord chosen perpendicular to the radius).

\begin{figure}
	\includegraphics[width=\textwidth]{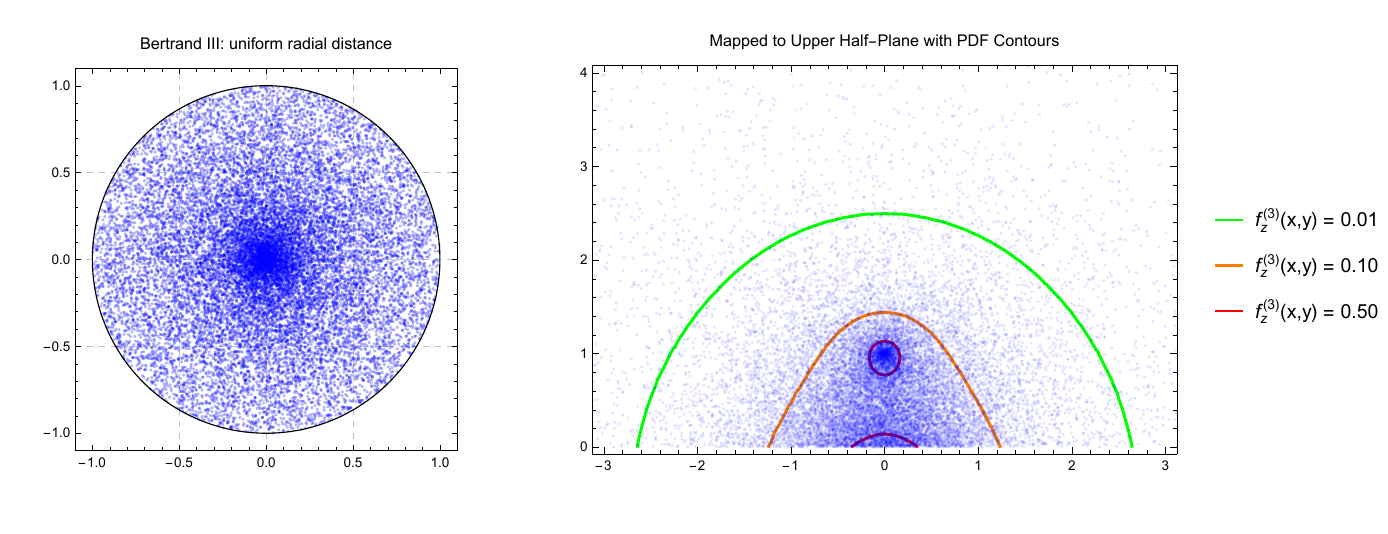}
	\caption{Bertrand~III (random radius midpoint) sampling, $N = 20{,}000$.}\label{fig:bertrandIII}
\end{figure}

To obtain the induced disk density, note that the joint density in polar coordinates is $\frac{1}{2\pi}$ on $ \theta \in[0,2\pi]$, while the area element is $du\,dv=r\,dr\,d\theta$. 
Hence 
\begin{equation}
	\,f_{\mathbb{D}}^{\text{III}}(u,v)
	=\frac{1}{2\pi\,\sqrt{u^2+v^2}} \,.
\end{equation}
The transformation~\eqref{eq:transformation-Jacobian} yields the probability density on  the upper half plane as
\begin{equation}
	\,f_{\mathbb{H}}^{\text{III}}(x,y)=\frac{2}{\pi}\,\frac{1}{\sqrt{x^2+(y-1)^2}\,\big(x^2+(y+1)^2\big)^{\frac{3}{2}}} \,.
\end{equation}
The resulting distribution is shown in Figure~\ref{fig:bertrandIII}.

\subsection{Algorithm and training}
\label{ssec:moduli-Algorithm}
Having sampled points according to the four different measures, we now proceed to test whether each point can be accurately mapped to its corresponding image within the fundamental domain.
This is accomplished by computing an SL$(2,\mathbb{Z})$ matrix that relates the original point to its image. A correct identification of the matrix is indicated by an accurate correspondence between the image and the original point. 
The data generating algorithm we employ is outlined in Algorithm~\ref{alg:fundamental_domain_reduction}.

Among several publicly available sequence-to-sequence Transformer architectures evaluated in our preliminary experiments, including the vanilla Transformer, BART, and T5, the T5 family consistently achieved the best validation accuracy and training stability. We therefore adopted the T5 family~\cite{Raffel2019ExploringTL} as the backbone throughout this work.
Depending on the complexity of the symbolic simplification task, different checkpoints within the T5 family were employed in the subsequent experiments.
Our implementation is based on the Hugging Face \texttt{transformers} library~\cite{Wolf2019HuggingFacesTS} with a PyTorch backend~\cite{Paszke2019PyTorchAI}. 
Unless noted otherwise, all model parameters were left at their default settings. 
Training was performed using the AdamW optimizer~\cite{loshchilov2017decoupled} with an initial learning rate of $3\times10^{-4}$, a batch size of $1024$, and the cross-entropy loss. 
The model converged after approximately $57$ epochs. All experiments were conducted on a single NVIDIA RTX 4060Ti GPU and required roughly $6$ hours of training.

\begin{algorithm}
	\caption{Fundamental Domain Reduction}\label{alg:fundamental_domain_reduction}
	\begin{algorithmic}[1]
		\Procedure{ReduceToFundamentalDomain}{$z = x + iy$}
		\State $M \gets I_2$
		\While{$|x| \geq \frac{1}{2}$}
		\State $n \gets \text{round}(x)$
		\State $M \gets M \cdot T_n$, $x \gets x - n$ \Comment{Translation by $T_n = \begin{pmatrix} 1 & -n \\ 0 & 1 \end{pmatrix}$}
		\EndWhile
		\While{$|z|^2 < 1$}
		\State $M \gets M \cdot S$, $z \gets S(z)$ \Comment{Inversion by $S = \begin{pmatrix} 0 & -1 \\ 1 & 0 \end{pmatrix}$}
		\While{$|\text{Re}(z)| \geq \frac{1}{2}$}
		\State Apply translation as above
		\EndWhile
		\EndWhile
		\State \Return $(z, M)$
		\EndProcedure
		
		\Procedure{GenerateDataset}{$N$}
		\For{$i = 1$ to $N$}
		\State Sample $z \notin \mathcal{F}$
		\State $(z', M) \gets \text{ReduceToFundamentalDomain}(z)$
		\State Store $(z, M)$ in dataset
		\EndFor
		\EndProcedure
	\end{algorithmic}
\end{algorithm}

The dataset comprised $1{,}000{,}000$ points, generated equally from four sampling methods: the hyperbolic measure (truncation radius $R_h=2$) and the Bertrand~I, II, and III schemes.
All data points were rounded to five decimal places and randomly partitioned into training (90\%) and validation (10\%) sets.
For evaluation, we generated separate test sets of $10{,}000$ points for each sampling method to ensure a consistent benchmark.
The quantitative results with five-decimal precision are summarized in Table~\ref{tab:results}.
The model achieves an average accuracy of \textbf{93.9\%} across all test sets.
\begin{table}[H]
	\centering
	\begin{tabular}{lc}
		\hline 
		\textbf{Test Set} & \textbf{Accuracy} \\ 
		\hline 
		Hyperbolic Measure ($R_h = 2$)  & 96.6\%\\
		Bertrand~I (random endpoints)    &  89.7\%\\
		Bertrand~II (uniform midpoint)   &  94.3\%\\
		Bertrand~III (uniform radial distance)   &  95.1\%\\
		\hline
	\end{tabular}
	\caption{Model accuracy on different test sets with five-decimal precision.}\label{tab:results}
\end{table}

The reduced accuracy for the Bertrand~I dataset is associated with the concentration of probability mass near the real axis (see Figure~\ref{fig:chordal_dist}), a feature shared by the hyperbolic measure with a large truncation radius (\textit{e.g.}, $R_h=10$).
In this regime, the small imaginary part increases the numerical sensitivity of the sampled modular parameters.
To test whether finite numerical precision is responsible for the observed loss of accuracy, we increased the numerical resolution from five to ten decimal places.
The resulting accuracies show no systematic improvement across the different sampling schemes, indicating that the numerical resolution is not the dominant limiting factor. 
The reduced performance is therefore more likely associated with the difficulty of sampling and learning the near-boundary region itself.
Notably, the model exhibits robustness by maintaining $>70\%$ accuracy even for $R_h=10$.

Our results therefore confirm with high confidence that the algorithm successfully recognizes M\"obius transformations in a numerical setting. In the following section, we will apply such SL$(2,\mathbb{Z})$  transformations to functions with modular properties, and investigate how symbolic expressions depending on the modulus $\tau$ simplify under these transformations.
Specific model configurations used for the individual datasets will be described in the corresponding sections.

\section{Machine learning modular functions}
\label{sec:ML-modularfunction}

In this section, we investigate the use of a machine learning algorithm to simplify formulas involving SL$(2,\mathbb{Z})$ transformations of the $q$-$\theta$ function and SL$(3,\mathbb{Z})$ transformations of the elliptic Gamma function $\Gamma(z;\tau,\sigma)$. Both functions arise as partition functions in four-dimensional superconformal field theories (SCFTs). For instance, the partition function of a vector multiplet on $S^3\times S^1$ is given by $\theta(z;\tau)$, while that of an $\mathcal{N}=1$ chiral multiplet on the same manifold involves elliptic Gamma functions. Theories with multiple chiral multiplets and richer flavor symmetries lead to complicated combinations of elliptic Gamma functions. Furthermore, theories defined on non-trivial backgrounds such as lens spaces $L(p,q)\times S^1$~\cite{Jejjala:2022lrm,Nieri:2015yia} also produce intricate combinations of these special functions.

Recall that the simplification of polylogarithm functions can reveal the analytic structure of scattering amplitude singularities, which contain crucial information about mass-shell conditions and propagators. Previous work~\cite{Dersy:2022bym} successfully employed machine learning techniques to handle the complexity of simplifying polylogarithmic expressions with high accuracy. Since multiple elliptic Gamma functions --- including $\theta(z;\tau)$ and $\Gamma(z;\tau,\sigma)$ --- appear as SCFT partition functions, their singularities are essential for understanding possible non-perturbative saddles in dual gravitational theories. Moreover, the locations of these singularities determine key features, such as the asymptotic growth of state degeneracies (see \textit{e.g.},~\cite{pemantle2024analytic,whittaker2020course}).
Modular transformations are beyond framework of the polylogarithm identities~\cite{Dersy:2022bym} studied by machine learning techniques. 
Therefore, whether machine learning can effectively predict modular transformations and utilize them to simplify complicated expressions requires additional methodological development.

Following the philosophy of Ref.~\cite{Dersy:2022bym}, the simplification of expressions involving multiple polylogarithms is formulated so as to minimize the number of polylogarithmic terms. To this end, a cost function is introduced, defined as the total count of polylogarithm functions appearing during the simplification process. 
Other components of the identity—specifically, the logarithmic and constant parts in \eqref{eq:dilog-identities}—do not incur additional cost, and the algorithm is accordingly biased toward reducing the polylogarithm count.
The remaining pieces of the identities, including logarithmic and constant terms, were not themselves treated as machine-learning targets, but were instead fixed by the underlying analytic identities, such as the pentagon identity.
In this paper, we adopt a similar philosophy: our primary objective is to simplify expressions by reducing the number of $q$-$\theta$ and elliptic Gamma functions throughout the procedure.
As in earlier machine-learning studies of polylogarithm identities, the learning task is formulated as the identification of a canonical representative. The modular cocycle is not learned separately, since it can be reconstructed deterministically by the algorithm presented in Appendix~\ref{app:phase}.

There is, however, one important distinction from the polylogarithm case. 
Under $S$-transformation, or more generally, under SL$(r,\mathbb{Z})$, the transformation of elliptic Gamma functions introduces a phase.
In the $q\to 1$ limit these phases reduce to the remainder terms appearing in the dilogarithm identities \eqref{eq:dilog-identities}. From the perspective of superconformal indices, such phases give the dominant contributions to the computation of black hole entropies and other root of unity saddle points \cite{Benini:2018ywd,Choi:2018hmj,Cabo-Bizet:2018ehj,Goldstein:2020yvj}. 
Nevertheless, once both the transformed canonical expression and the original expression are known, the modular cocycle can be reconstructed deterministically (up to an additive constant) by iteratively applying the difference equations of the elliptic Gamma function \cite{Felder_2000,Jejjala:2022lrm}.
Accordingly, the machine learning task considered in this section is the identification of canonical representatives involving only the $q$-$\theta$ and elliptic Gamma functions, while the modular cocycle is reconstructed separately by the deterministic procedure described in Appendix~\ref{app:phase}.

\subsection{Machine learning for $q$-$\theta$ function}

\subsubsection{Data preparation}
\label{ssec:data-theta}
The identities to generate a general transformation on $\theta(z;\tau)$ functions (including SL$(2,\mathbb{Z})$ modular transformation) involve (see~\cite{Jejjala:2022lrm} for complete review of these identities):
\begin{align}\label{eq:transq --- extra}
	\begin{array}{rclcl}
	\theta(z;\tau) & \rightarrow & \theta(z;\tau +1) & \qquad & \texttt{(T-transformation)} \,, \\
	\theta(z;\tau) & \rightarrow & \theta\left(\frac{z}{\tau};-\frac{1}{\tau}\right) & & \texttt{(S-transformation)} \,, \\
	\theta(z;\tau) &\rightarrow & \theta(z+\tau;\tau) & & \texttt{(shift)} \,, \\
	\theta(z;\tau) &\rightarrow & \theta(\tau - z; \tau) & & \texttt{(reflection)} \,, \\
	\theta(z;\tau) &\rightarrow & 1/\theta(z; -\tau) & & \texttt{(inversion)} \,, \\
	\theta(z,\tau) &\rightarrow & \prod_{a,b \in \{0,1\}} \theta \left( \frac{z + a\tau + b}{2}; \tau \right) & & \texttt{(duplication)} \,, \\
	\theta(z;\tau) & \rightarrow & \theta(z;\tau) & & \texttt{(identity)} \,.
	\end{array}
\end{align}
A subset of the transformations in~\eqref{eq:transq --- extra} includes the generators $T,S$ and the identity, which together span the full SL$(2,\mathbb{Z})$ modular group. Any generic SL$(2,\mathbb{Z})$ transformation can be decomposed into a product of $n_s$ factors of $S,T$ matrices; the minimal number of factors required --- the word length --- serves as a measure of its generation cost. More general transformations within the set~\eqref{eq:transq --- extra} will be denoted by $\mathcal{M}$. 
Our objective is to simplify a complicated expression into the following target canonical form:
\begin{equation} \label{eq:qsimple}
	\frac{\prod_{i=1}^{i_{max}}\theta(\frac{z}{c_i \tau +d_i};\frac{a_i \tau +b_i}{c_i \tau +d_i})}{\prod_{j=1}^{j_{max}}\theta(\frac{z}{c_j \tau +d_j};\frac{a_j \tau +b_j}{c_j \tau +d_j})}\,,
\end{equation} 
where the integers $i_{max}$ and $j_{max}$ respectively limit the numbers of $q$-$\theta$ functions in the numerator and the denominator, and no further cancellation is possible.
This expression can be viewed as a vector of length $i_{max}+j_{max}$, analogous to a quantum state in a Hilbert space:
\begin{equation}
	|\Theta\rangle = \left( \bigotimes_{i=1}^{i_{max}} |f_i\rangle \otimes \bigotimes_{j=1}^{j_{max}} |f_j^{-1}\rangle \right)\,,
\end{equation}
where each single-factor state $ |f\rangle$  takes the value $|1 \rangle$ if the corresponding $\theta$ expression is nontrivial, and  $|0 \rangle$ if the  $\theta$-function cancels to be $1$.

To evaluate the model's ability to predict simplifications under modular transformations, we consider two settings for the transformation set $\mathcal{M}$: one restricted to modular transformations alone, and another allowing all permissible transformations. 
Given an SL$(2,\mathbb{Z})$ matrix defining a modular transformation, its action on a $q$-$\theta$ function is defined as
\begin{equation}
	\bigl(\mathcal{A}\circ\theta\bigr)(z;\tau):=\theta\bigl(\mathcal{A}\cdot (z;\tau)\bigr)\,.
\end{equation}
Data are generated by applying several SL$(2,\mathbb{Z})$ matrices $\mathcal{A}_i,\mathcal{B}_j,\mathcal{C}_u,\mathcal{D}_u$ to the irreducible form~\eqref{eq:qsimple}, producing scrambled expressions of the form:
\begin{equation}
	\frac{\prod_{i=1}^{i_{max}}\bigl(\mathcal{A}_{i}\circ\theta\bigr)\!\left(\frac{z}{c_i \tau +d_i};\frac{a_i \tau +b_i}{c_i \tau +d_i}\right)}{\prod_{j=1}^{j_{max}}\bigl(\mathcal{B}_{j}\circ\theta\bigr)\!\left(\frac{z}{c_j \tau +d_j};\frac{a_j \tau +b_j}{c_j \tau +d_j}\right)}\times \prod_{u=1}^{n_t}\frac{\bigl(\mathcal{C}_{u}\circ\theta\bigr)(z;\tau)}{\bigl(\mathcal{D}_{u}\circ\theta\bigr)(z;\tau)} \,,
\end{equation}
where $n_t$ counts additional $\theta$ function pairs that can be simplified via the modular identity~\eqref{eq:thetaSL2Z}. 
The matrices $\mathcal{C}_u,\mathcal{D}_u$ serve as scrambling transformations that extend the sequence length, increasing the dimension of the $\theta$-function vector space to $(i_{max}+j_{max}+2n_t)$:
\begin{equation}\label{eq:state-theta-ini}
	|\Theta_{\text{init}}\rangle = \left( \bigotimes_{i=1}^{i_{max}} |f_i\rangle \otimes \bigotimes_{j=1}^{j_{max}} |f_j^{-1} \rangle \right) \otimes \left( \bigotimes_{s=1}^{n_t} \underbrace{|f_s \rangle \otimes |f_s^{-1} \rangle}_{\text{trivial pairs}} \right).
\end{equation}
Within computational limits, we fix $n_s,n_t$ and $i_{max},j_{max}$ as specified in Table~\ref{tab:da-qtheta} and generate a total of 500,000 data points.
\begin{table}[htbp]
	\centering
	\begin{tabular}{|c|c|}
		\hline
		\begin{tabular}[c]{@{}c@{}}
			\text{Complicated expressions} \\
			\parbox{11cm}{\centering
				\begin{flalign*}
					\frac{\prod_{i=1}^{i_{max}}\bigl(\mathcal{A}_{i}\circ\theta\bigr)\!\left(\frac{z}{c_i \tau +d_i};\frac{a_i \tau +b_i}{c_i \tau +d_i}\right)}{\prod_{j=1}^{j_{max}}\bigl(\mathcal{B}_{j}\circ\theta\bigr)\!\left(\frac{z}{c_j \tau +d_j};\frac{a_j \tau +b_j}{c_j \tau +d_j}\right)}\times \prod_{u=1}^{n_t}\frac{\bigl(\mathcal{C}_{u}\circ\theta\bigr)(z;\tau)}{\bigl(\mathcal{D}_{u}\circ\theta\bigr)(z;\tau)}
				\end{flalign*}
			} \\[2ex]
			\hline
			\text{Simplified expressions} \\
			\parbox{11cm}{\centering
				\begin{flalign*}
					\frac{\prod_{i=1}^{i_{max}}\theta(\frac{z}{c_i \tau +d_i};\frac{a_i \tau +b_i}{c_i \tau +d_i})}{\prod_{j=1}^{j_{max}}\theta(\frac{z}{c_j \tau +d_j};\frac{a_j \tau +b_j}{c_j \tau +d_j})}
				\end{flalign*}
			}
		\end{tabular}
		&
		\parbox[c]{3.25cm}{\centering
			Random parameters \\[10pt]
			$n_{s} \in [3, 5]$ \\
			$n_{t} \in [0, 2]$ \\
			$i_{max} +j_{max} \in [1,3]$
		} \\
		\hline
	\end{tabular}
	\caption{Structure of the input and output data.}\label{tab:da-qtheta}
\end{table}

For more intricate expressions, actions from set $\mathcal{M}$ can also alter the number of $\theta(z;\tau)$ functions via duplication identities~\eqref{eq:duplication}. 
The scrambling operator $\mathcal{M}$  is defined as an ordered sequence of $n_s$ elementary operators drawn from~\eqref{eq:transq --- extra}, \textit{i.e.}, $\mathcal{M} = \mathcal{O}_{n_s} \circ \dots \circ \mathcal{O}_1$. 
Each operator $\mathcal{O}_k$ the formula space and is constructed as a tensor product of a single local non‑trivial transformation $\hat{T}$ (randomly chosen from the generating set~\eqref{eq:transq --- extra}) with identity operators $\mathbbm{1}$ on all other factors
\begin{equation}\label{eq:compositeoperator-theta}
	\mathcal{O}_k = \cdots \otimes \mathbbm{1} \otimes \mathbbm{1}\otimes \cdots \otimes \underbrace{\hat{T}_{\text{elem}}}_{\mathclap{\text{selected factor}}} \otimes \cdots \otimes \mathbbm{1} \otimes \cdots,
\end{equation}
Applying such operators to a state of the form~\eqref{eq:state-theta-ini} typically produces a new state; simplification occurs when the total number of $q$-$\theta$ functions is reduced. 
These actions are systematically summarized in Table~\ref{tab:da-qtheta-extra}.
\begin{table}[htbp]
	\centering
	\begin{tabular}{|c|c|}
\hline
\begin{tabular}[c]{@{}c@{}}
\text{Complicated expressions} \\
\parbox{11cm}{\centering
	\begin{flalign*}
\mathcal{M} \circ \left( \frac{\prod_{i=1}^{i_{max}} \theta \left(\frac{z}{c_i \tau +d_i};\frac{a_i \tau +b_i}{c_i \tau +d_i}\right)}{\prod_{j=1}^{j_{max}} \theta \left(\frac{z}{c_j \tau +d_j};\frac{a_j \tau +b_j}{c_j \tau +d_j}\right)} \times \prod_{u=1}^{n_t}\frac{\theta(z;\tau)}{\theta(z;\tau)} \right)
	\end{flalign*}
	} \\[2ex]
	\hline
	\text{Simplified expressions} \\
\parbox{11cm}{\centering
\begin{flalign*}
\frac{\prod_{i=1}^{i_{max}}\theta(\frac{z}{c_i \tau +d_i};\frac{a_i \tau +b_i}{c_i \tau +d_i})}{\prod_{j=1}^{j_{max}}\theta(\frac{z}{c_j \tau +d_j};\frac{a_j \tau +b_j}{c_j \tau +d_j})}
				\end{flalign*}
			}
		\end{tabular}
		&
		\parbox[c]{3.25cm}{\centering
			Random parameters \\[10pt]
			$n_{s} \in [3, 5]$ \\
			$n_{t} \in [0, 2]$ \\
			$i_{max} +j_{max} \in [1,3]$ 
		} \\
		\hline
	\end{tabular}
	\caption{Structure of $q$-$\theta$ dataset with additional transformations.}
	\label{tab:da-qtheta-extra}
\end{table}

We emphasize that the forward scrambling procedure used to generate the dataset is distinct from the inverse simplification problem. 
While a complicated expression can be generated deterministically by applying a recorded sequence of identities to a canonical expression, the simplification task receives only the final expression, without access to its scrambling history. 
The model is therefore required to recover the canonical representative directly from the scrambled expression.

The generated data consist of analytical expressions, requiring a preprocessing step to encode these expressions into suitable matrix representations in order to be processed by Neural networks.
Mathematical expressions can be modeled as tree structures and serialized into \textbf{prefix notation}, yielding a more compact token sequence. 
Subsequently, each token is mapped to a unique identifier via a constructed dictionary, thereby completing the transformation from symbolic expressions to matrix-compatible format~\cite{DBLP:journals/corr/abs-1912-01412}.
The following figure demonstrates the conversion process of the expression $\log(x+1) - 2x^3 + 7$ from its standard form to the matrix input format.

\begin{minipage}[t]{0.3\textwidth}
	\centering
	\textbf{\small Expression Tree}\\[2em]
	\Tree[.$+$ [.$7$ ] [.$-$ [.$\log$ [.$+$ [.$x$ ][.$1$ ]]] [.$\times$ [.$2$ ][.$\mathrm{pow}$ [.$x$ ][.$3$ ]]]]]
\end{minipage}%
\hfill
\begin{minipage}[t]{0.3\textwidth}
	\centering
	\textbf{\small Prefix notation}\\
	\begin{equation*}
		\begin{pmatrix}
	+ \\ 7 \\ - \\ \log \\ + \\ x \\ 1 \\ \times \\ 2 \\ \mathrm{pow} \\ x \\ 3
		\end{pmatrix}
	\end{equation*}
\end{minipage}%
\hfill
\begin{minipage}[t]{0.3\textwidth}
	\centering
	\textbf{\small Token ID sequence}\\
	\begin{equation*}
		\begin{pmatrix}
7 \\ 21 \\ 8 \\ 19 \\ 7 \\ 1 \\ 18 \\ 9 \\ 13 \\ 12 \\ 1 \\ 11
		\end{pmatrix}
	\end{equation*}
\end{minipage}

\subsubsection{Training and verification}
\label{ssec:training-theta}

Our objective is to identify canonical representatives of expressions involving
$\theta(z;\tau)$ functions, while the modular cocycle in~\eqref{eq:thetaSL2Z}
is reconstructed separately using the deterministic procedure described in
Appendix~\ref{app:phase}.
Following the backbone selection strategy described in Sec \ref{ssec:moduli-Algorithm},
we employ the
Flan-T5-small checkpoint for  the present symbolic simplification task. 
Preliminary comparisons with other representative sequence-to-sequence Transformer architectures are
summarized in Table~\ref{tab:qtheta-backbone}, motivating the choice of the Flan-T5-small checkpoint.
\begin{table}[H]
	\centering
\setlength{\tabcolsep}{15pt}
    \begin{tabular}{lcc}
        \hline
        \textbf{Backbone} & \textbf{Parameters} & \textbf{Validation Accuracy} \\
        \hline
        Vanilla Transformer & $\sim 6\times 10^ 7$ & $\sim$10\% \\
        BART-base & $\sim 1.39\times 10^8$ & $\sim$30\% \\
        T5-small & $\sim 6 \times 10^7 $ & $\sim$95\% \\
        Flan-T5-small & $\sim 7.7\times 10^7$ & $\sim$99\% \\
        \hline
    \end{tabular}
	\caption{Preliminary backbone comparison on the $q$-$\theta$ symbolic simplification task. 
The reported validation accuracies are approximate results from exploratory experiments.
T5-small and Flan-T5-small achieve substantially higher validation accuracy than the Vanilla Transformer and BART-base. 
In particular, the larger parameter count of BART-base does not lead to better performance in this task.
	}
	\label{tab:qtheta-backbone}
\end{table}

The simplification is framed as a sequence-to-sequence task: the model is trained in a supervised manner to reproduce target expressions obtained from exact symbolic reduction. Training minimizes the cross-entropy loss, which is equivalent to maximizing the log-likelihood of the target token sequence given the model's predicted distribution.
Because the target sequences are canonical simplified forms with redundant terms removed, minimizing the cross-entropy loss encourages the model to reproduce these canonical representations faithfully.
Consequently, the fine-tuned Flan-T5 model is trained to map scrambled expressions to their corresponding canonical symbolic representations.

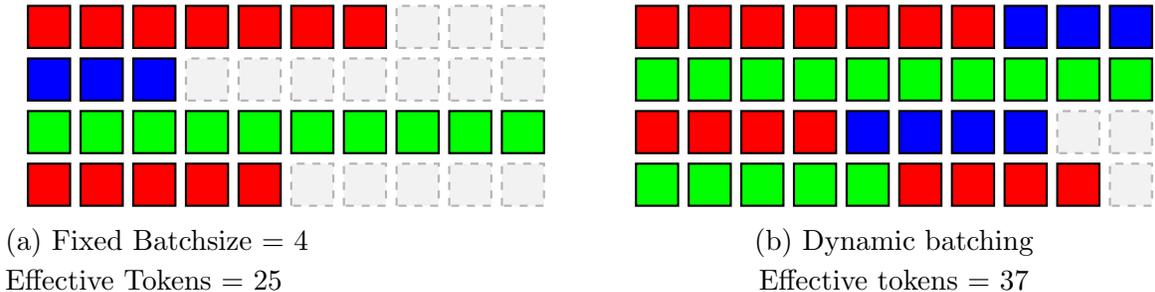
\begin{figure}[H]
\centering

\tikzset{
  block/.style={
    draw=black, 
    thick,       
    rectangle    
  },
  padding/.style={
    draw=gray!60,
    thick,
    dashed,
    rectangle,
    fill=gray!10
  }
}

\begin{subfigure}[b]{0.48\textwidth}
\centering
\begin{tikzpicture}
    [x=0.7cm, y=0.7cm]

    \foreach \x in {0,...,6} {
      \fill[red, block] (\x, 3) rectangle ++(0.8, 0.8);
    }
    \foreach \x in {7,...,9} {
      \draw[padding] (\x, 3) rectangle ++(0.8, 0.8);
    }

    \foreach \x in {0,...,2} { 
      \fill[blue, block] (\x, 2) rectangle ++(0.8, 0.8);
    }
    \foreach \x in {3,...,9} {
      \draw[padding] (\x, 2) rectangle ++(0.8, 0.8);
    }

    \foreach \x in {0,...,9} { 
      \fill[green, block] (\x, 1) rectangle ++(0.8, 0.8);
    }

    \foreach \x in {0,...,4} {
      \fill[red, block] (\x, 0) rectangle ++(0.8, 0.8);
    }
    \foreach \x in {5,...,9} {
      \draw[padding] (\x, 0) rectangle ++(0.8, 0.8);
    }
\end{tikzpicture}
\caption{Fixed Batchsize = 4 \\ Effective Tokens = 25}
\label{fig:seq_len_8}
\end{subfigure}
\hfill 
\begin{subfigure}[b]{0.48\textwidth}
\centering
\begin{tikzpicture}
    [x=0.7cm, y=0.7cm]

    \foreach \x in {0,...,6} {
      \fill[red, block] (\x, 3) rectangle ++(0.8, 0.8);
    }
    \foreach \x in {7,...,9} {
      \fill[blue, block] (\x, 3) rectangle ++(0.8, 0.8);
    }

    \foreach \x in {0,...,9} {
      \fill[green, block] (\x, 2) rectangle ++(0.8, 0.8);
    }

    \foreach \x in {0,...,3} {
      \fill[red, block] (\x, 1) rectangle ++(0.8, 0.8);
    }
    \foreach \x in {4,...,7} {
      \fill[blue, block] (\x, 1) rectangle ++(0.8, 0.8);
    }
    \foreach \x in {8,...,9} {
      \draw[padding] (\x, 1) rectangle ++(0.8, 0.8);
    }

    \foreach \x in {0,...,4} {
      \fill[green, block] (\x, 0) rectangle ++(0.8, 0.8);
    }
    \foreach \x in {5,...,8} {
      \fill[red, block] (\x, 0) rectangle ++(0.8, 0.8);
    }
    \draw[padding] (9, 0) rectangle ++(0.8, 0.8);
\end{tikzpicture}
\caption{Dynamic batching \\ Effective tokens = 37} 
\label{fig:seq_len_12}
\end{subfigure}

\caption{
Comparison between fixed-size batching and dynamic batching under the same total token budget. 
Gray dashed blocks indicate padding. Dynamic batching minimizes padding overhead, thereby accommodating more effective tokens (colored blocks) within the same budget.
}
\label{fig:com_dyn}
\end{figure}

To reduce the computational cost of handling variable‑length sequences, we introduce a Dynamic Batching Algorithm. Instead of fixing the number of sequences per batch, our method constrains the total number of tokens per batch by a fixed token budget (token size). Sequences are first sorted by length and then grouped greedily so that the cumulative token count in a batch stays below a predefined threshold. This approach avoids the excessive padding required by conventional fixed‑length batching, which is especially wasteful when sequence lengths vary widely. As shown in Figure~\ref{fig:com_dyn}, dynamic batching leads to more efficient use of computational resources.
In practice, we observe that this strategy yields a \textbf{30\%} reduction in training time compared to fixed‑size batching.

Following a similar methodology, each mathematical expression is processed through a multi‑stage encoding pipeline. First, the expression is parsed into a syntax tree, which is then traversed to produce a token sequence in prefix notation. In the end, this token sequence is mapped one‑to‑one into a matrix representation.

Following the comparison described in Sec \ref{ssec:moduli-Algorithm}, we employ the same Flan-T5-small backbone for symbolic simplification.
The model was trained on a single NVIDIA RTX 5090 GPU with a token size of 25600. Training concluded after approximately 3 hours, with the model converging around epoch 36. This efficiency is largely attributed to the Dynamic Batching Algorithm. To avoid unusually long sequences, we cap the token sequence length at $L_{max}=512$ during preprocessing and discard samples whose length exceeds $L_{max}$.

\begin{table}[H]
	\begin{center}
		\renewcommand{\arraystretch}{2}  
		\begin{tabular}{|c|}
			\hline
			\text{Complicated expressions} \\ 
			\parbox{14cm}{\centering
				$\displaystyle
				\frac{
\theta\!\left(\frac{-z}{3\tau-1};\frac{5\tau-2}{3\tau-1}\right)
\theta\!\left(\frac{z}{3\tau+2};\frac{-4\tau-3}{3\tau+2}\right)
 \theta\!\left(\frac{z}{6\tau+11};\frac{5\tau+9}{6\tau+11}\right)
 \theta\!\left(\frac{z}{16\tau-7};\frac{9\tau-4}{16\tau-7}\right)
}{
\theta\!\left(\frac{z}{\tau+1};\frac{6\tau+5}{\tau+1}\right)
\theta\!\left(\frac{-z}{3\tau+2};\frac{11\tau+7}{3\tau+2}\right)
\theta\!\left(\frac{z}{5\tau-2};\frac{1-3\tau}{5\tau-2}\right)^2
\theta\!\left(\frac{-z}{6\tau-1};\frac{11\tau-2}{6\tau-1}\right)
}$ } \\[0.5cm]  
			\hline
\text{Simplified expressions} \\  
			\parbox{14cm}{\centering
				$\displaystyle
				\frac{\theta\!\left(\frac{-z}{\tau+2};\frac{4\tau+7}{\tau+2}\right)}
				{\theta\!\left(\frac{-z}{3\tau+2};\frac{2\tau+1}{3\tau+2}\right)
				\theta\!\left(\frac{-z}{5\tau-1};\frac{1-6\tau}{5\tau-1}\right)}$ } \\[0.3cm]  
			\hline
		\end{tabular}
		\caption{Sample expressions from the simplification dataset of $q$-$\theta$ function.}\label{tab:qtheta-example}
	\end{center}
\end{table}

\subsubsection*{Numerical Verification}

In this study, our model is trained to simplify complicated products of $\theta(z;\tau)$ functions into compact standard forms. A fundamental challenge in verifying the model's predictions arises from the non-trivial transformation properties of these functions. Under the action of the modular group $\mathrm{SL}(2,\mathbb{Z})$, the $\theta(z;\tau)$ functions are invariant up to an exponential phase factor, which are second order diagonal Bernoulli polynomials~\eqref{eq:theB-polynomial}.
Consequently, a predicted simplification $f_{\text{pred}}(z,\tau)$ is considered correct if it relates to the input expression $f_{\text{input}}(z,\tau)$ by such an allowed phase.
For the large test sets we use the following numerical check to test whether the logarithmic ratio has the expected quadratic dependence on $z$.
This check decides whether the prediction lies in the allowed phase class; for an accepted input-output pair the relative phase can be reconstructed algebraically, as described in Appendix~\ref{app:qtheta-phase}.
We compare the partial derivatives on the $\ln(\frac{f_{\text{pred}}}{f_{\text{input}}})$ and verify numerically whether it is a linear function of $z$ following the procedure below:
\begin{enumerate}
	\item \textbf{Random Sampling:} For each test sample, we randomly generate a modular parameter $\tau$ with $\mathrm{Im}(\tau) > 0$ to avoid singular boundaries. 
	We then sample $N=10$ random points $\{z_k\}_{k=1}^N$ uniformly in the domain $z_k \in [-0.5, 0.5] \times [-0.5, 0.5]i$.
	
	\item \textbf{Numerical Differentiation:} We compute the logarithmic derivative for both the input and predicted expressions using the central difference method. For a small step size $h=10^{-5}$, the approximation is given by:
	\begin{equation}
	\partial\ln (f)(z_k) \approx \frac{	 f(z_k+h) - 	f(z_k-h)}{2h \cdot 	f(z_k)}.
	\end{equation}
	This formulation avoids the ambiguity of the complex logarithm function by computing the ratio of the derivative to the function value directly.
	
	\item \textbf{Linearity Test:} We calculate the set of $y_k= \partial_z \ln(\frac{f_{\text{pred}}}{f_{\text{input}}})$ and perform a complex linear least-squares fit to the model linear in $z$. 
	The prediction is accepted as correct if the mean squared residual of the fit satisfies:
	\begin{equation}
		\frac{1}{N} \sum_{k=1}^{N} \left| y_k - (\hat{\alpha} z_k + \hat{\beta}) \right|^2 < \epsilon,
	\end{equation}
	where $\epsilon = 10^{-3}$ is set as the tolerance threshold and $\hat{\alpha},\hat{\beta}$ are fitting coefficients which are $\tau$-dependent.
\end{enumerate}
Our codes are provided on Github \cite{fan_jejjala_lei_ML_Modularity}.

\subsubsection{Training outcomes}

Building on the verification framework described above, we generated an additional 10,000 test samples using the same parameter ranges as the training set: $n_{s} \in [3, 5]$, $n_{t} \in [0, 2]$ and $i_{max} +j_{max} \in [1,3]$. 
We achieve following results.
\begin{itemize}
	\item The model to simplify formula achieved an accuracy of \textbf{99.96\%} on the in‑distribution test set if the transformation only involves SL$(2,\mathbb{Z})$ type transformation.
	\item To further evaluate generalization beyond the training distribution, we extended the parameter $n_s$ to the range $[6,10]$ and generated another 10,000 samples. 
	Even on this more challenging set beyond the training inputs, the model retained a high accuracy of \textbf{99.87\%} with only  SL$(2,\mathbb{Z})$ type transformation being considered.
	\item The model to simplify formula achieved an accuracy of over \textbf{91\%} on the in‑distribution test set if the transformation involves all the kinds of transformations in~\eqref{eq:transq --- extra}.
\end{itemize}
The decreasing of accuracy rate in the model with all the transformations is possibly due to the increasing length of the formula, which are hard to be captured by the program trained by shorter sequence of formula. 
These results confirm its robust capacity to capture the structural rules of composite modular transformations.

\subsection{Machine learning for elliptic Gamma functions}

\subsubsection{Data preparation}
The identities of elliptic Gamma functions input in the training is generated by the following sets of actions: including shifts in $\mathbb{Z}^3$, SL$(3,\mathbb{Z})$ modularity, inversions~\cite{Felder_2000}, and duplications~\cite{felder2002multiplication}.
These are
\begin{equation}
\begin{array}{rclcl}
	\Gamma (z;\tau,\sigma) & \rightarrow & \Gamma(z;\sigma,\tau) & \qquad & \texttt{(symmetry)} \,, \\
	\Gamma(z;\tau,\sigma) & \rightarrow & \Gamma(z+1;\tau,\sigma) && \texttt{(periodicity-z)} \,, \\
	\Gamma(z;\tau,\sigma) & \rightarrow & \Gamma(z;\tau +1,\sigma) && \texttt{(periodicity-$\tau$)} \,, \\
	\Gamma(z;\tau,\sigma) & \rightarrow & \Gamma(z;\tau,\sigma+1) && \texttt{(periodicity-$\sigma$)} \,, \\
	\Gamma(z;\tau,\sigma) & \rightarrow & \frac{1}{\Gamma(\tau+\sigma-z;\tau,\sigma)} && \texttt{(inversion)} \,, \\
	\Gamma(z;\tau,\sigma) & \rightarrow & \frac{1}{\Gamma(z-\tau;-\tau,\sigma)} && \texttt{(shift-1)} \,, \\
	\Gamma(z;\tau,\sigma) & \rightarrow & \Gamma(\sigma -z;-\tau,\sigma) && \texttt{(shift-2)} \,, \\
	\Gamma(z;\tau,\sigma) & \rightarrow & \Gamma(z;\tau-\sigma,\sigma)\Gamma(z;\sigma-\tau,\tau) && \texttt{(mod-1)} \,, \\
	\Gamma(z;\tau,\sigma) & \rightarrow & \Gamma\left(\frac{z}{\sigma};\frac{\tau}{\sigma},-\frac{1}{\sigma}\right)\Gamma\left(\frac{z}{\tau};\frac{\sigma}{\tau},-\frac{1}{\tau} \right) && \texttt{(mod-2)} \,, \\
     \Gamma(z,\tau,\sigma) & \rightarrow & \prod_{a,b,c \in \{0,1\}}\Gamma\left(\tfrac{z+a \tau +b \sigma +c}{2}, \tau, \sigma\right) && \texttt{(duplication)} \,, \\
	\Gamma(z;\tau,\sigma) & \rightarrow & \Gamma(z;\tau,\sigma) && \texttt{(identity)} \,.\label{eq:action1}
\end{array}
\end{equation}
To avoid heavy notation, we introduce the following convention for the elliptic Gamma function under the action of the SL$(3,\mathbb{Z})$ modular group:
\begin{align}
\begin{split}
	f_{i}^{(\sigma)}(z;\tau,\sigma) &=\Gamma\left(\frac{z}{m_i \sigma + n_i}; \frac{\tau - \tilde{n}_{i}(k_i \sigma + l_i)}{m_i \sigma + n_i}, \frac{k_i \sigma + l_i}{m_i \sigma + n_i}\right) \,, \\
	f_{i}^{(\tau)} (z;\tau,\sigma) &=\Gamma\left(\frac{z}{m_i \tau + \tilde{n}_i}; \frac{\sigma - n_i(\tilde{k}_i \tau + \tilde{l}_i)}{m_i \tau + \tilde{n}_i}, \frac{\tilde{k}_i \tau + \tilde{l}_i}{m_i \tau + \tilde{n}_i}\right)\label{eq:general-SL3Z-M} \,.
	\end{split}
\end{align}
We then aim to simplify complicated expressions involving elliptic Gamma functions of the following form:
\begin{align}
	\frac{\prod_{i=1}^{i_{max}}f_{i}^{(u)}(z;\tau,\sigma)}{\prod_{j=1}^{j_{max}}f_{j}^{(v)}(z;\tau,\sigma)}\label{eq:simple}\,,
\end{align}
where $u,v\in\{\tau,\sigma\}$ are randomly chosen.
Together with $n_t$ cancellable pairs being introduced, the initial expressions span the vector space  of the state 
\begin{equation}
	|\Psi_{\text{init}}\rangle = \left( \bigotimes_{i=1}^{i_{max}} |f_i^{(u)}\rangle \otimes \bigotimes_{j=1}^{j_{max}} |f_j^{(v)}\rangle^{-1} \right) \otimes \left( \bigotimes_{s=1}^{n_t} \underbrace{|f_s^{(w)}\rangle \otimes |f_s^{(w)}\rangle^{-1}}_{\text{trivial pairs}} \right).
\end{equation}

For each \( f \), we apply \( n_s \) elementary transformations from~\eqref{eq:action1} and compose them into a single composite operator  $\mathcal{M}$ made by the composite operator~\eqref{eq:compositeoperator-theta}. 
Compared to the $\theta$-function cases, more transformations including \texttt{mod-1}, \texttt{mod-2} and \texttt{dup} can map a single elliptic Gamma function to a product of elliptic Gamma functions, potentially increasing the number of factors in the expression. 
We then apply the operator $\mathcal{M}$ acting on the seed expressions to generate a scrambled initial expression: 
\begin{flalign*}
	\mathcal{M} \circ \left( \frac{\prod_{i=1}^{i_{max}}f_{i}^{(u)}(z;\tau,\sigma)}{\prod_{j=1}^{j_{max}}f_{j}^{(v)}(z;\tau,\sigma)}\times \prod_{s=1}^{n_{t}}\frac{f_{s}^{(w)}(z;\tau,\sigma)}{f_{s}^{(w)}(z;\tau,\sigma)} \right) \,.
\end{flalign*}
We summarize the data structure used in the simplification task of elliptic Gamma function in Table~\ref{tab:egam}.
\begin{table}[H]
	\begin{center}
		\footnotesize
		\renewcommand{\arraystretch}{1.55}
		\begin{tabular}{|c|}
			\hline
			\text{Complicated expressions} \\
			\parbox{14cm}{\centering
				\[
				\resizebox{0.97\linewidth}{!}{%
				$\displaystyle
				\begin{aligned}
				&
				\Gamma\!\left(-\frac{2z}{4\tau+7};\frac{-\sigma-28\tau-49}{4\tau+7},\frac{3\tau+5}{4\tau+7}\right)
				\Gamma\!\left(\frac{2z}{\sigma+28\tau+49};\frac{-4\tau-7}{\sigma+28\tau+49},\frac{-3\tau-5}{\sigma+28\tau+49}\right)\\
				&\quad\times
				\Gamma\!\left(\frac{\sigma+2z-8}{\sigma-8};\frac{\tau+3}{\sigma-8},\frac{\sigma-9}{\sigma-8}\right)
				\Bigl[
				\Gamma\!\left(-\frac{z}{2\tau+6};\frac{3\tau+8}{\tau+3},\frac{-\sigma-18\tau-48}{\tau+3}\right)\\
				&\quad\times
				\Gamma\!\left(\frac{2z}{3\tau+5};\frac{\sigma+28\tau+49}{3\tau+5},\frac{-4\tau-7}{3\tau+5}\right)
				\Gamma\!\left(\frac{\tau-z+3}{2\tau+6};\frac{3\tau+8}{\tau+3},\frac{-\sigma-18\tau-48}{\tau+3}\right)\\
				&\quad\times
				\Gamma\!\left(\frac{3\tau-z+8}{2\tau+6};\frac{3\tau+8}{\tau+3},\frac{-\sigma-18\tau-48}{\tau+3}\right)
				\Gamma\!\left(\frac{4\tau-z+11}{2\tau+6};\frac{3\tau+8}{\tau+3},\frac{-\sigma-18\tau-48}{\tau+3}\right)\\
				&\quad\times
				\Gamma\!\left(\frac{-\sigma-18\tau-z-48}{2\tau+6};\frac{3\tau+8}{\tau+3},\frac{-\sigma-18\tau-48}{\tau+3}\right)
				\Gamma\!\left(\frac{-\sigma-17\tau-z-45}{2\tau+6};\frac{3\tau+8}{\tau+3},\frac{-\sigma-18\tau-48}{\tau+3}\right)\\
				&\quad\times
				\Gamma\!\left(\frac{-\sigma-15\tau-z-40}{2\tau+6};\frac{3\tau+8}{\tau+3},\frac{-\sigma-18\tau-48}{\tau+3}\right)
				\Gamma\!\left(\frac{-\sigma-14\tau-z-37}{2\tau+6};\frac{3\tau+8}{\tau+3},\frac{-\sigma-18\tau-48}{\tau+3}\right)
				\Bigr]^{-1}
				\end{aligned}
				$}
				\]
			} \\[0.2cm]
			\hline
			\text{Simplified expressions} \\
			\parbox{14cm}{\centering
				\[
				\Gamma\!\left(\frac{2z}{\sigma-8};\frac{\tau+3}{\sigma-8},-\frac{1}{\sigma-8}\right)
				\Bigl[
				\Gamma\!\left(-\frac{z}{\tau+3};\frac{-\sigma-18\tau-48}{\tau+3},\frac{3\tau+8}{\tau+3}\right)
				\Bigr]^{-1}.
				\]
			} \\[0.2cm]
			\hline
		\end{tabular}
		\caption{Sample expressions from the simplification dataset of elliptic Gamma functions.}\label{tab:egam}
	\end{center}
\end{table}

\subsubsection{Training and numerical tests}

Compared with the $q$-$\theta$ function, the additional moduli in elliptic Gamma functions introduce substantially more elementary transformations, significantly increasing the complexity of the symbolic mapping task. 
Accordingly, we employ the larger Flan-T5-base checkpoint~\cite{chung2022scalinginstructionfinetunedlanguagemodels}, which belongs to the same T5 family introduced in Sec.~\ref{ssec:moduli-Algorithm}. 
The larger model capacity was chosen to accommodate the increased complexity of the symbolic simplification problem. Moreover, to accommodate the longer scrambled expressions, the maximum encoder sequence length was increased to 1024 tokens, and samples exceeding this limit were discarded.

Training was performed on a single NVIDIA RTX 5090 GPU. We used a token size of 8192 with 4 gradient‑accumulation steps, which helped stabilize training. The dynamic batching strategy described in Section~\ref{ssec:training-theta} was also employed to improve computational efficiency. 
Under this setup, the model converged quickly, reaching optimal performance in about 15 epochs.
The entire training process required roughly 30 hours. 

The training and test sets were generated independently using different random seeds. Since the symbolic expressions are constructed from random compositions over a finite discrete parameter space, exact overlap between the two sets cannot be ruled out a priori. 
We therefore additionally evaluate the model on extrapolation tasks involving parameter regimes that are deliberately excluded from the training data.

To verify the simplified expressions numerically, we adopted a method analogous to that used for the $q$-$\theta$ functions in Section~\ref{ssec:training-theta}. 
This large-scale check tests whether the input and predicted expressions differ by an allowed Bernoulli phase.
As in the theta case, the numerical criterion is used for evaluation, while the actual phase can be recovered by a deterministic certificate in the space of Gamma products; this reconstruction is given in Appendix~\ref{app:egamma-phase}.
Because the phase is the sum of a set of third order Bernoulli polynomial in $z$,  the logarithmic ratio between the input and predicted expressions, denoted $\ln(\frac{f_{\text{pred}}}{f_{\text{input}}})$, is expected to follow a cubic dependence in $z$.
Verification proceeds by evaluating this ratio function for a set of random points $z_k = z_0 + k h$ and test whether its fourth finite difference vanishes:
\begin{equation}
	\Delta^4 [\ln R(z)] = \sum_{k=0}^4 (-1)^k \binom{4}{k} \ln R_k = 0 \,,
	\quad \quad 
	\mathcal{I} \equiv \frac{R_0 R_4 R_2^6}{R_1^4 R_3^4} = 1 \,,
\end{equation}
where $R_k = R(z_k)$. 
A prediction is accepted as correct if $|\mathcal{I} - 1| < 10^{-3}$. Codes are provided in Github also \cite{fan_jejjala_lei_ML_Modularity}.

\subsubsection{Results}

To rigorously evaluate the model's performance and robustness, we carried out a two‑stage testing procedure. First, we constructed a test set of 50,000 samples drawn uniformly from the  \textit{interpolation regime} ($n_s \in [3,6]$, $n_t \in [0,2]$), which matches the parameter distribution of the training data. On this in‑distribution set, the model attained an accuracy of \textbf{97.23\%}. 
Subsequently, to assess the model's extrapolation capability, we generated an additional 50,000 samples from an out‑of‑distribution parameter space where $n_s \in [7,10]$. 
Even on this more challenging generalization set, the model retained a robust accuracy of \textbf{93.02\%}.

\begin{figure}[t]
	\centering
	\includegraphics[width=0.7\linewidth]{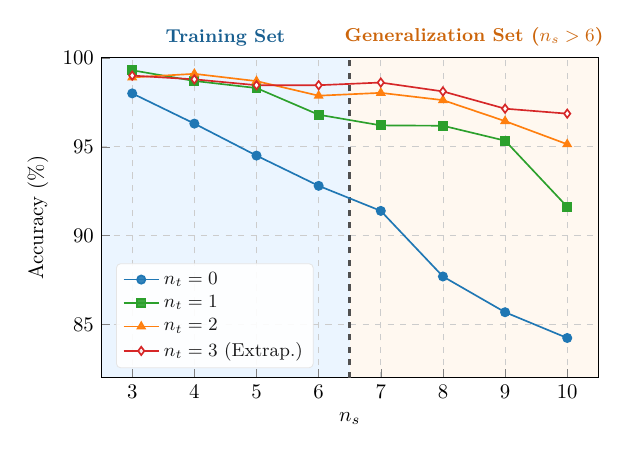} 
	\caption{Prediction accuracy as a function of scrambling depth $n_s$ and the number of identity insertions $n_t$. The model is trained only on $n_s \in [3,6]$ and $n_t=0,1,2$. Results for $n_s\in[7,10]$ and $n_t=3$ correspond to extrapolation settings excluded from the training data.}
	\label{fig:egamma_acc}
\end{figure}

To disentangle the specific influence of scrambling depth ($n_s$) and the number of identity insertions ($n_t$) on performance, we performed a systematic grid scan. 
The model was trained only on samples with $n_s \in [3,6]$ and $n_t=0,1,2$. 
For each parameter pair $(n_s,n_t)$, we then created an independent evaluation set of 5,000 samples, covering both the interpolation regime and extrapolation regimes beyond the training parameter range, including samples with $n_s\in[7,10]$ and the ones with $n_t=3$. 
The resulting scaling behavior is shown in Figure~\ref{fig:egamma_acc}, where the results are obtained using a beam size of $2$.
To examine whether beam search improves the performance at larger scrambling depth,
we further tested beam sizes $1,2,5,7,$ and $10$ on 2,000 random samples at
$(n_s,n_t)=(10,3)$, corresponding to one of the most challenging extrapolation settings.
The verified accuracies remain within a narrow range of approximately $96\%$--$98\%$
and show no systematic improvement as the beam size is increased.
In contrast, the inference and verification time grows substantially, from about
$300$ seconds for beam size $1$ to approximately $2200$ seconds for beam size $10$.
These results indicate that increasing the beam size provides little benefit for the
present task while introducing a considerable computational cost.

As seen in the figure, prediction accuracy decreases monotonically with increasing $n_s$.
This is expected because $n_s$ directly quantifies the entropy introduced by scrambling operations, making the simplification task more difficult. 
In contrast, accuracy increases with the number of identity insertions $n_t$. We attribute this to a decrease in the effective scrambling density --- \textit{i.e.}, the ratio of scrambling operations to the total length of the expression.
As $n_t$ increases, more identity (redundant) terms are inserted; for a fixed $n_s$, the scrambling operations thus become more sparsely distributed across the expression. This dilution of scrambling effects makes it easier for the model to recognize and cancel the inserted identities, leading to higher accuracy even in the extrapolation regime ($n_t=3$).

In summary, our results demonstrate the model's robust ability to identify and simplify complex identities involving elliptic Gamma functions.
Even under deep scrambling sequences, the model maintains high prediction accuracy, achieving over \textbf{95\%} in key extrapolation scenarios.
Its strong performance on previously unseen parameter regimes --- specifically for scrambling depths ($n_s>6$) and an unseen number of identity insertions ($n_t=3$) --- indicates that the learned mapping generalizes well beyond the symbolic expressions encountered during training.

A natural physical application of the present framework arises in the modular factorization of supersymmetric lens indices on $L(p,q)\times S^1$ \cite{Jejjala:2022lrm}. 
These indices are expressed as products of elliptic Gamma functions and satisfy identities of the form 
\begin{eqnarray} \nonumber
	& &\left( \prod_{i=0}^{t-1}\Gamma (z+ p_{i-1}\tau-s_{i-1}\sigma;  p_i \tau- s_i \sigma\,, p_{i-1}\tau-s_{i-1} \sigma)\right) \Gamma (z;  p_t \tau- s_t \sigma\,, p_{t-1}\tau-s_{t-1} \sigma)\\ \label{eq:mod-prop-Gamma}
	&=&e^{-i\pi \tilde{P}_{g_{(p,q)}}^{\mathbf{m}}(z,\tau,\sigma)}\Gamma\left(\tfrac{z}{m\sigma+n_1}; \tfrac{\tau-c(k_1\sigma+l_1)}{m\sigma+n_1}, \tfrac{k_1\sigma +l_1}{m\sigma+n_1} \right) \\
	\nonumber
	&&\qquad \qquad \qquad \times  \Gamma\left(\tfrac{z}{m(p\tau
		-s \sigma) +{\tilde{n}}_{t+1}}; \tfrac{q\tau-r\sigma-n_{t+1}(\tilde{k}_{t+1} (p \tau-s \sigma) +\tilde{l}_{t+1})}{m(p\tau-s \sigma) +{\tilde{n}}_{t+1}} , \tfrac{\tilde{k}_{t+1} (p \tau-s \sigma) +\tilde{l}_{t+1}}{m(p\tau-s\sigma) +{\tilde{n}}_{t+1}}\right)\,,
\end{eqnarray}
where the lens space is defined by a pair of coprime integers $(p_t,q_t) \equiv (p,q)$ admitting the continued fraction expansion
\begin{align}\label{eq:poveruandpoverq}
	\begin{split}
		\frac{p_i}{q_i}&=	 [e_i,e_{i-1},\ldots,e_{1}]^-\equiv e_i-\cfrac{1}{e_{i-1}-\frac{1}{\cdots-\frac{1}{e_1}}}\,,
	\end{split} 
\end{align}
and the sequences $p_i,q_i,s_i,r_i$ are determined by the recursive relations and SL$(2,\mathbb{Z})$ relation $s_i q_i- p_i r_i=1$ as
\begin{equation}\label{eq:recursive-pq}
	\begin{aligned}
		&	p_{i} = e_{i} p_{i-1}- p_{i-2}\,, \qquad   \, q_i=p_{i-1}\,, \\
		&s_{i} = e_{i} s_{i-1}- s_{i-2} \,,\qquad   r_i=s_{i-1}\,,\\
		& p_0=1\,,\quad p_1= e_1 \,, \quad	s_0=0\,, \quad s_1 =1 \,.
	\end{aligned}
\end{equation}
The parameters $n_i,k_i,l_i,m$ specify a general SL$(3,\mathbb{Z})$ transformation subject to $k_1n_1-l_1m=1$ and $
\tilde{k}_{t+1}\tilde{n}_{t+1}-\tilde{l}_{t+1}m=1$, with
\begin{equation}
	n_{t+1}=qc-rn_1\,,\qquad \tilde{n}_{t+1}=-sn_1+pc\,. 
\end{equation}
Finally the $\tilde{P}$ in \eqref{eq:mod-prop-Gamma} is a polynomial that depends on the geometric data and deforms the third-order diagonal Bernoulli polynomial.

The modular factorization of lens indices is obtained by successively applying the
SL$(3,\mathbb Z)$ modular transformation of the elliptic Gamma function.
The \eqref{eq:mod-prop-Gamma} illustrates the typical symbolic complexity encountered during
the modular factorization procedure.
Although each elementary transformation is simple, repeated applications rapidly
generate lengthy intermediate expressions involving products of elliptic Gamma functions with shifted arguments. Consequently, modular factorization becomes,
to a large extent, a problem of symbolic simplification.

The machine-learning framework developed in the present work is designed precisely for this task. 
Given a complicated expression generated by repeated modular
transformations, the model identifies an equivalent canonical representative containing with reduced elliptic Gamma function content. 
The associated modular cocycle can then be reconstructed deterministically using the algorithm presented in Appendix~\ref{app:phase}, thereby recovering the complete modular transformation.
Looking beyond lens spaces, similar modular identities also arise for partition functions on more general Seifert three-manifolds $M_3\times S^1$ \cite{Closset:2018ghr},
whose modular structures remain much less understood. 
We expect that the present framework may provide a useful symbolic tool for investigating these more general
factorization problems.

\section{Discussion}

In this work, we present a machine learning framework for simplifying formulas containing $q$-$\theta$ and elliptic Gamma functions by directly utilizing their associated identities, including the full SL$(r,\mathbb{Z})$ modular transformations. 
Our results demonstrate that the proposed framework can reproduce algebraic simplifications based on these modular identities. 
This task goes beyond merely predicting numerical attributes, such as the weight of a modular form from its Fourier coefficients~\cite{Jejjala:2025hgv}, by directly performing symbolic transformations between equivalent expressions.

The natural extensions of this work proceed in two directions. 
First, to advance towards practical applications --- such as building a simplification package in Mathematica or similar systems --- it is essential to understand how the model behaves when expressions contain both $q$-$\theta$ and elliptic Gamma functions $\Gamma(z;\tau,\sigma)$, also combined with polylogarithm functions.
For example, the function $T(z;\tau)$ defined in~\eqref{eq:T=thetaLi2} serves as a bridge between these three classes of functions which transform under the SL$(2,\mathbb{Z})$ modular group~\cite{Pa_ol_2018}. 
Extending our current framework to incorporate such hybrid expressions would necessitate the introduction of a higher-spin quantum state representation, where the basis of states $|0\rangle$, $\cdots$, $|s \rangle$ correspond respectively to phase factors and the distinct function types under study. 
This generalization would substantially increase computational complexity and therefore merits a dedicated investigation.

The second direction is to extend the framework to handle formulas involving higher-rank multiple elliptic Gamma functions and multiple sine functions~\cite{Nishizawa_2001,kurokawa2003multiple}, whose modular properties have recently been shown to govern the partition functions of free scalar conformal field theories~\cite{Lei:2024oij}. Importantly, however, partition functions of other physically relevant systems --- such as free fermion or Maxwell theories \cite{Basar:2015asd,Basar:2015xda} --- do not belong to the class of multiple elliptic Gamma functions. This observation underscores the broader importance of investigating identity relations for more general families of elliptic functions.

\section*{Acknowledgements}

We thank Arghya Chattopadhyay, Sanjaye Ramgoolam, Andrew Turner, Chi Zhang, and Xiaoyuan Zhang for useful discussions.
We also thank the anonymous referee for helpful comments that improved the manuscript.
V.J.\ thanks the Institute of Advanced Study, Soochow University for hospitality when this work was in progress.
Y.F.\ is supported by Undergraduate Training Program for Innovation and Entrepreneurship, Soochow University (Grant No.\ 202410285037Z).
V.J.\ is supported by the South African Research Chairs Initiative of the Department of
Science, Technology, and Innovation (DSTI) and the National Research Foundation (NRF), grant 78554.
V.J.\ is grateful to the ``50 years of the black hole information paradox'' program at the Simons Center for Geometry and Physics at Stony Brook University and the ``Generative AI for high and low energy physics'' program at the Kavli Institute for Theoretical Physics at the University of California, Santa Barbara, the latter of which was supported in part by grant NSF PHY-2309135, for their hospitality.
Y.L.\ is supported by a Project Funded by the Priority Academic Program Development of Jiangsu Higher Education Institutions (PAPD) and by the National Natural Science Foundation of China (NSFC) No.12305081 and the international collaboration and communication grant between NSFC and the Royal Society No.W2421035.

\appendix

\section{Direct reconstruction of the phase }
\label{app:phase}

\subsection{$q$-$\theta$ phase}
\label{app:qtheta-phase}
For the $q$-$\theta$ dataset, once the complicated expression and the simplified expression are both given, the relative phase can be reconstructed directly from the two expressions. In this appendix we summarize the reconstruction algorithm and work out explicitly the example displayed in Table~\ref{tab:qtheta-example}.

\subsubsection{General procedure}

Each theta factor appearing in the dataset is first put in the normalized form
\begin{equation}
	\theta\!\left(\frac{z}{c\tau+d};\frac{a\tau+b}{c\tau+d}\right),
	\qquad
	\begin{pmatrix}
		a & b \\
		c & d
	\end{pmatrix}
	\in \mathrm{SL}(2,\mathbb{Z}).
\end{equation}

All phases below are understood modulo $2$, since only the factor $e^{i\pi Q}$ appears in the transformation formula.
For practical applications, it is convenient to work with the generators
\begin{equation}
	T=
	\begin{pmatrix}
		1 & 1 \\
		0 & 1
	\end{pmatrix},
	\qquad
	S=
	\begin{pmatrix}
		0 & -1 \\
		1 & 0
	\end{pmatrix},
\end{equation}
together with the corresponding elementary transformation laws
\begin{align}
	\theta(z;\tau+n) &= \theta(z;\tau), \qquad n\in \mathbb{Z}, \label{eq:app-theta-T} \\
	\theta\!\left(\frac{z}{\tau};-\frac{1}{\tau}\right)
	&=
	e^{i\pi B(z;\tau)}\theta(z;\tau), \label{eq:app-theta-S}
\end{align}
where $B(z;\tau)$ is related to the second diagonal Bernoulli polynomial, explicitly as
\begin{equation}
	B(z;\tau)=\frac{z^2}{\tau}+z\left(\frac{1}{\tau}-1\right)+\frac{1}{6}\left(\tau+\frac{1}{\tau}\right)-\frac{1}{2}.
\end{equation}

The reconstruction algorithm proceeds as follows.
\begin{enumerate}
	\item For each theta factor, identify the normalized modular matrix
	\(
	A=\begin{psmallmatrix}a&b\\ c&d\end{psmallmatrix}
	\)
from the transformed argument
	\(
	\theta\bigl(\frac{z}{c\tau+d};\frac{a\tau+b}{c\tau+d}\bigr)
	\).
	\item Decompose $A$ into a word in $S$ and $T^n$ using Euclid algorithm. If $c=0$, then $A$ reduces to a set of $T$ words. Otherwise, 
	\begin{equation}
		q=\mathrm{round}\!\left(\frac{a}{c}\right),
		\qquad
		A'=ST^{-q}A.
	\end{equation}
	where round$(x)$ denotes the nearest integer to $x$.
Since $|c'|<|c|$, the procedure terminates after finitely many iterations.
\item Starting from $(z,\tau)$, evaluate the resulting word from left to right. 
Each occurrence of $S$ contributes a phase $B(z;\tau)$ and updates
\begin{equation}
		(z;\tau)\longmapsto \left(\frac{z}{\tau};-\frac{1}{\tau}\right),
	\end{equation}
	while every $T^n$ contributes no phase and updates
	\begin{equation}
		(z,\tau)\longmapsto (z,\tau+n).
	\end{equation}
The total cocycle phase is obtained by summing all $B(z;\tau)$ contributions 
	\begin{equation}
Q=Q_{\mathrm{cocycle}} = \sum_k B(z_k;\tau_k)\,.
\end{equation}
	\item For a product of $\theta$ functions
	\(
	F=\prod_j [\theta(\cdots)]^{p_j}
	\),
	the total phase is 
	\begin{equation}
		Q_{\mathrm{total}}=\sum_j p_j Q_j,
	\end{equation}
	where $p_j=+1$ for numerator factors and $p_j=-1$ for denominator factors.
	\item For two equivalent expressions $F_{\mathrm{complicated}}$ and $F_{\mathrm{simple}}$,
	\begin{equation}
		F_{\mathrm{complicated}}=e^{i\pi Q_{\mathrm{rel}}}F_{\mathrm{simple}},
		\qquad
		Q_{\mathrm{rel}}=Q_{\mathrm{scr}}-Q_{\mathrm{can}}.
	\end{equation}
\end{enumerate}

\subsubsection{Worked example corresponding to Table~\ref{tab:qtheta-example}}

We now apply the above procedure to the example used in Table~\ref{tab:qtheta-example}. After the normalization above, the simplified expression is
\begin{equation}
	F_{\mathrm{simple}}
	=
	\frac{\theta\!\left(\frac{z}{-\tau-2};\frac{-4\tau-7}{-\tau-2}\right)}
	{\theta\!\left(\frac{z}{-3\tau-2};\frac{-2\tau-1}{-3\tau-2}\right)
	\theta\!\left(\frac{z}{-5\tau+1};\frac{6\tau-1}{-5\tau+1}\right)},
	\label{eq:app-Fcan}
\end{equation}
while the corresponding complicated expression is
\begin{equation}
	F_{\mathrm{complicated}}
	=
	\frac{
	\theta\!\left(\frac{z}{-3\tau+1};\frac{-5\tau+2}{-3\tau+1}\right)
	\theta\!\left(\frac{z}{3\tau+2};\frac{-4\tau-3}{3\tau+2}\right)
	\theta\!\left(\frac{z}{6\tau+11};\frac{5\tau+9}{6\tau+11}\right)
	\theta\!\left(\frac{z}{16\tau-7};\frac{9\tau-4}{16\tau-7}\right)
	}{
	\theta\!\left(\frac{z}{\tau+1};\frac{6\tau+5}{\tau+1}\right)
	\theta\!\left(\frac{z}{-3\tau-2};\frac{-11\tau-7}{-3\tau-2}\right)
	\theta\!\left(\frac{z}{5\tau-2};\frac{1-3\tau}{5\tau-2}\right)^2
	\theta\!\left(\frac{z}{-6\tau+1};\frac{-11\tau+2}{-6\tau+1}\right)
	},
	\label{eq:app-Fscr}
\end{equation}

\subsubsection*{Matrix extraction}

For every factor
\(
\theta\bigl(\frac{z}{c\tau+d};\frac{a\tau+b}{c\tau+d}\bigr)
\)
we extract the matrix
\(
\begin{psmallmatrix}a&b\\ c&d\end{psmallmatrix}
\),
and the power $p$.

\begin{table}
	\centering
	\renewcommand{\arraystretch}{2.0}
	\begin{tabular}{ccc}
		\toprule
		Factor & Matrix & $p$ \\
		\midrule
		$\theta\!\left(\frac{z}{-\tau-2};\frac{-4\tau-7}{-\tau-2}\right)$ & $\begin{psmallmatrix}-4&-7\\ -1&-2\end{psmallmatrix}$ & $+1$ \\
		$\theta\!\left(\frac{z}{-3\tau-2};\frac{-2\tau-1}{-3\tau-2}\right)$ & $\begin{psmallmatrix}-2&-1\\ -3&-2\end{psmallmatrix}$ & $-1$ \\
		$\theta\!\left(\frac{z}{-5\tau+1};\frac{6\tau-1}{-5\tau+1}\right)$ & $\begin{psmallmatrix}6&-1\\ -5&1\end{psmallmatrix}$ & $-1$ \\
		\bottomrule
	\end{tabular}
	\caption{Matrices for the factors in the canonical expression~\eqref{eq:app-Fcan}.}
\end{table}

\begin{table}
	\centering
	\small
	\setlength{\tabcolsep}{8pt}
	\renewcommand{\arraystretch}{2.0}
	\begin{tabular}{cccccc}
		\toprule
		Factor & Matrix & $p$
		& Factor & Matrix & $p$ \\
		\midrule
		$\theta\!\left(\frac{z}{-3\tau+1};\frac{-5\tau+2}{-3\tau+1}\right)$ & $\begin{psmallmatrix}-5&2\\ -3&1\end{psmallmatrix}$ & $+1$
		& $\theta\!\left(\frac{z}{-3\tau-2};\frac{-11\tau-7}{-3\tau-2}\right)$ & $\begin{psmallmatrix}-11&-7\\ -3&-2\end{psmallmatrix}$ & $-1$ \\
		$\theta\!\left(\frac{z}{3\tau+2};\frac{-4\tau-3}{3\tau+2}\right)$ & $\begin{psmallmatrix}-4&-3\\ 3&2\end{psmallmatrix}$ & $+1$
		& $\theta\!\left(\frac{z}{5\tau-2};\frac{1-3\tau}{5\tau-2}\right)$ & $\begin{psmallmatrix}-3&1\\ 5&-2\end{psmallmatrix}$ & $-1$ \\
		$\theta\!\left(\frac{z}{6\tau+11};\frac{5\tau+9}{6\tau+11}\right)$ & $\begin{psmallmatrix}5&9\\ 6&11\end{psmallmatrix}$ & $+1$
		& $\theta\!\left(\frac{z}{5\tau-2};\frac{2\tau-1}{5\tau-2}\right)$ & $\begin{psmallmatrix}2&-1\\ 5&-2\end{psmallmatrix}$ & $-1$ \\
		$\theta\!\left(\frac{z}{16\tau-7};\frac{9\tau-4}{16\tau-7}\right)$ & $\begin{psmallmatrix}9&-4\\ 16&-7\end{psmallmatrix}$ & $+1$
		& $\theta\!\left(\frac{z}{-6\tau+1};\frac{-11\tau+2}{-6\tau+1}\right)$ & $\begin{psmallmatrix}-11&2\\ -6&1\end{psmallmatrix}$ & $-1$ \\
		$\theta\!\left(\frac{z}{\tau+1};\frac{6\tau+5}{\tau+1}\right)$ & $\begin{psmallmatrix}6&5\\ 1&1\end{psmallmatrix}$ & $-1$
		& \multicolumn{3}{c}{} \\
		\bottomrule
	\end{tabular}
	\caption{Matrices for the factors in the complicated expression~\eqref{eq:app-Fscr}.}
\end{table}

\subsubsection*{Detailed decomposition for the first canonical factor}

Consider the factor
\begin{equation}
	\theta\!\left(\frac{z}{-\tau-2};\frac{-4\tau-7}{-\tau-2}\right),
	\qquad
	A_1=
	\begin{pmatrix}
		-4 & -7 \\
		-1 & -2
	\end{pmatrix}.
\end{equation}
We decompose $A_1$ by the Euclidean algorithm.

\paragraph{Recursion round 1.}
Since $c=-1\neq 0$, we compute
\begin{equation}
	q=\mathrm{round}\!\left(\frac{a}{c}\right)=\mathrm{round}(4)=4.
\end{equation}
Then
\begin{equation}
	T^{-4}A_1
	=
	\begin{pmatrix}
		1 & -4 \\
		0 & 1
	\end{pmatrix}
	\begin{pmatrix}
		-4 & -7 \\
		-1 & -2
	\end{pmatrix}
	=
	\begin{pmatrix}
		0 & 1 \\
		-1 & -2
	\end{pmatrix},
\end{equation}
and therefore
\begin{equation}
	A_1'
	=
	ST^{-4}A_1
	=
	\begin{pmatrix}
		0 & -1 \\
		1 & 0
	\end{pmatrix}
	\begin{pmatrix}
		0 & 1 \\
		-1 & -2
	\end{pmatrix}
	=
	\begin{pmatrix}
		1 & 2 \\
		0 & 1
	\end{pmatrix}.
\end{equation}
The new lower-left entry is $c'=0$, so the recursion terminates.

\paragraph{Base case and backtracking.}
Since
\(
A_1'=\begin{psmallmatrix}1&2\\ 0&1\end{psmallmatrix}
\)
has $c=0$, we may write
\begin{equation}
	A_1' = T^2.
\end{equation}
Consequently,  the word returned by the recursion is
\begin{equation}
	W_1'=[T^2].
\end{equation}
Using
\(
A_1'=ST^{-4}A_1
\),
we obtain
\(
A_1=T^4S^{-1}A_1'=T^4S^3A_1'
\),
so the full word is
\begin{equation}
	W_1=[T^2,S,S,S,T^4].
\end{equation}
Equivalently, we have
\begin{equation}
	A_1=T^4\cdot S^3\cdot T^2
	=
	\begin{pmatrix}
		-4 & -7 \\
		-1 & -2
	\end{pmatrix}.
\end{equation}

\subsubsection*{Cocycle accumulation for the first canonical factor}

Starting from $(z,\tau)$, we apply the word $W_1=[T^2,S,S,S,T^4]$ from left to right.

\paragraph{Step 1.}
The transformation $T^2$ contributes no phase and maps
\begin{equation}
	(z,\tau)\longmapsto (z,\tau+2).
\end{equation}

\paragraph{Steps 2 and 3.}
The first two $S$-transformations then contribute
\begin{align}
	B(z;\tau+2)
	&=
	\frac{z^2}{\tau+2}+\frac{z}{\tau+2}-z+\frac{\tau}{6}-\frac{1}{6}+\frac{1}{6(\tau+2)}, \\
	B\!\left(\frac{z}{\tau+2};-\frac{1}{\tau+2}\right)
	&=
	-\frac{z^2}{\tau+2}-z-\frac{z}{\tau+2}-\frac{\tau}{6}-\frac{5}{6}-\frac{1}{6(\tau+2)}.
\end{align}
Adding them gives
\begin{equation}
	B(z;\tau+2)+B\!\left(\frac{z}{\tau+2};-\frac{1}{\tau+2}\right)=-2z-1.
\end{equation}

\paragraph{Step 4.}
The last nontrivial generator is another $S$, with contribution
\begin{equation}
	B(-z;\tau+2)
	=
	\frac{z^2}{\tau+2}-\frac{z}{\tau+2}+z+\frac{\tau}{6}-\frac{1}{6}+\frac{1}{6(\tau+2)}.
\end{equation}

\paragraph{Step 5.}
The final $T^4$ again contributes no phase and sends
\begin{equation}
	\left(\frac{-z}{\tau+2},-\frac{1}{\tau+2}\right)
	\longmapsto
	\left(\frac{z}{-\tau-2},\frac{-4\tau-7}{-\tau-2}\right),
\end{equation}
which matches the target argument and modular parameter.

Collecting the nonzero contributions, the cocycle phase is
\begin{align}
	Q_1
	={}&
	(-2z-1)
	+
	\left(\frac{z^2}{\tau+2}-\frac{z}{\tau+2}+z+\frac{\tau}{6}-\frac{1}{6}+\frac{1}{6(\tau+2)}\right)
	\notag\\
	={}&
	\frac{\tau^2-6\tau z-5\tau+6z^2-18z-13}{6(\tau+2)}.
	\label{eq:app-Q1}
\end{align}

\subsubsection*{Phases of the remaining factors}

Applying the same procedure to the other two factors in~\eqref{eq:app-Fcan} gives
\begin{align}
	Q_2
	&=
	\frac{\tau^2-6\tau z-7\tau+6z^2-6z-5}{2(3\tau+2)}, \\
	Q_3
	&=
	\frac{5\tau^2-30\tau z-35\tau+30z^2+7}{6(5\tau-1)}.
\end{align}
Hence the total phase of the canonical expression is
\begin{align}
	Q_{\mathrm{can}}
	&=
	Q_1-Q_2-Q_3 \notag\\
	&=
	\frac{\tau^2-6\tau z-5\tau+6z^2-18z-13}{6(\tau+2)}
	-
	\frac{\tau^2-6\tau z-7\tau+6z^2-6z-5}{2(3\tau+2)}
	\notag\\
	&\quad
	-
	\frac{5\tau^2-30\tau z-35\tau+30z^2+7}{6(5\tau-1)} \notag\\
	&=
	\frac{
	(-15\tau^4+75\tau^3+277\tau^2+96\tau-32)
	}{
	6(15\tau^3+37\tau^2+12\tau-4)
	}
	\notag\\
	&\qquad
	+
	\frac{
	z^2(-90\tau^2-360\tau-96)
	+z(90\tau^3+180\tau^2+132\tau)
	}{
	6(15\tau^3+37\tau^2+12\tau-4)
	}.
	\label{eq:app-Qcan}
\end{align}

For the nine factors in the complicated expression~\eqref{eq:app-Fscr}, the resulting phases are summarized in Table~\ref{tab:qtheta-phase-scrambled}.

\begin{table}
	\centering
	\renewcommand{\arraystretch}{1.2}
	\resizebox{\textwidth}{!}{
	\begin{tabular}{|c|c|c|}
		\hline
	 $\begin{psmallmatrix}a&b\\ c&d\end{psmallmatrix}$ & $p$ & Phase $Q_j$ \\
		\hline
	 $\begin{psmallmatrix}-5&2\\ -3&1\end{psmallmatrix}$ & $+1$ &
		$\tfrac{\tau^2-6\tau z-9\tau+6z^2+3}{2(3\tau-1)}$ \\
	 $\begin{psmallmatrix}-4&-3\\ 3&2\end{psmallmatrix}$ & $+1$ &
		$\tfrac{\tau^2-6\tau z-13\tau+6z^2-2z-9}{2(3\tau+2)}$ \\
	 $\begin{psmallmatrix}5&9\\ 6&11\end{psmallmatrix}$ & $+1$ &
		$\tfrac{\tau^2-6\tau z-8\tau+6z^2-10z-18}{6\tau+11}$ \\
	 $\begin{psmallmatrix}9&-4\\ 16&-7\end{psmallmatrix}$ & $+1$ &
		$\tfrac{2(4\tau^2-24\tau z-164\tau+24z^2+12z+71)}{3(16\tau-7)}$ \\
	 $\begin{psmallmatrix}6&5\\ 1&1\end{psmallmatrix}$ & $-1$ &
		$\tfrac{\tau^2-6\tau z-13\tau+6z^2-13}{6(\tau+1)}$ \\
	 $\begin{psmallmatrix}-11&-7\\ -3&-2\end{psmallmatrix}$ & $-1$ &
		$\tfrac{\tau^2-6\tau z-7\tau+6z^2-6z-5}{2(3\tau+2)}$ \\
	 $\begin{psmallmatrix}-3&1\\ 5&-2\end{psmallmatrix}$, $\begin{psmallmatrix}2&-1\\ 5&-2\end{psmallmatrix}$ & $-1$ &
		$\tfrac{5\tau^2-30\tau z-199\tau+30z^2+18z+79}{6(5\tau-2)}$ each \\
	 $\begin{psmallmatrix}-11&2\\ -6&1\end{psmallmatrix}$ & $-1$ &
		$\tfrac{\tau^2-6\tau z-6\tau+6z^2+1}{6\tau-1}$ \\
		\hline
	\end{tabular}}
	\caption{Phases of the factors in the complicated expression~\eqref{eq:app-Fscr}.}
	\label{tab:qtheta-phase-scrambled}
\end{table}

Using the powers of the numerator and denominator factors, we obtain
\begin{equation}
	Q_{\mathrm{scr}}=\sum_{j=1}^{9} p_j Q_j^{(\mathrm{scr})}.
\end{equation}
The relative phase between the two expressions is therefore
\begin{equation}
	Q_{\mathrm{rel}}=Q_{\mathrm{scr}}-Q_{\mathrm{can}}.
\end{equation}
After simplification it takes the quadratic form
\begin{equation}
	Q_{\mathrm{rel}}(z,\tau)=\alpha_2(\tau)z^2+\alpha_1(\tau)z+\alpha_0(\tau),
\end{equation}
with
\begin{align}
	\alpha_2(\tau)
	&=
	\frac{6}{D(\tau)}
	\Bigl(
	65340\tau^7+103428\tau^6-79233\tau^5-103922\tau^4
	\notag\\
	&\qquad
	+106011\tau^3-34940\tau^2+6912\tau-776
	\Bigr), \\
	\alpha_1(\tau)
	&=
	-\frac{6}{D(\tau)}
	\Bigl(
	26460\tau^8+336024\tau^7+531633\tau^6-144118\tau^5
	\notag\\
	&\qquad
	-264393\tau^4+35312\tau^3+51810\tau^2-15748\tau+1200
	\Bigr), \\
	\alpha_0(\tau)
	&=
	\frac{1}{D(\tau)}
	\Bigl(
	272700\tau^9+845496\tau^8+310359\tau^7-724997\tau^6
	\notag\\
	&\qquad
	-121436\tau^5+221247\tau^4-12713\tau^3-22260\tau^2+5440\tau-376
	\Bigr),
\end{align}
where
\begin{equation}
	D(\tau)
	=
	6(\tau+1)(\tau+2)(3\tau-1)(3\tau+2)(5\tau-2)(5\tau-1)(6\tau-1)(6\tau+11)(16\tau-7).
\end{equation}
Thus the two expressions in Table~\ref{tab:qtheta-example} satisfy the exact relation
\begin{equation}
	F_{\mathrm{complicated}}=e^{i\pi Q_{\mathrm{rel}}(z,\tau)}F_{\mathrm{simple}}.
\end{equation}

\subsection{The phase for the elliptic Gamma function}
\label{app:egamma-phase}

For the elliptic Gamma dataset, once both the complicated and the simplified expressions are specified, the relative phase can likewise be reconstructed directly from their ratio. This is not an independent phase computation for elliptic Gamma functions. Instead, it relies on the difference equations of the elliptic Gamma function developed in Ref.~\cite{Felder_2000}, which express appropriate quotients of shifted Gamma functions as theta functions. 
As a result, the nontrivial cocycle of the elliptic Gamma function is reduced to the theta cocycle computed in Appendix~\ref{app:qtheta-phase}. The desired Gamma phase is then recovered by iteratively applying these difference equations.

\subsubsection{Gamma ratios and theta phases}

The starting point is the pair of difference equations
\begin{align}
	\frac{\Gamma(u+\sigma;\tau,\sigma)}{\Gamma(u;\tau,\sigma)}
	&=\theta(u;\tau),
	\label{eq:app-egamma-theta-shift-sigma}\\
	\frac{\Gamma(u+\tau;\tau,\sigma)}{\Gamma(u;\tau,\sigma)}
	&=\theta(u;\sigma),
	\label{eq:app-egamma-theta-shift-tau}
\end{align}
together with $\Gamma(u+1;\tau,\sigma)=\Gamma(u;\tau,\sigma)$. These identities are the mechanism that converts Gamma ratios into theta functions. In particular, the relevant modular Gamma ratio is
\begin{equation}
	R(u;\tau,\sigma)
	=
	\frac{
	\Gamma\!\left(\frac{u}{\sigma};\frac{\tau}{\sigma},-\frac{1}{\sigma}\right)
	\Gamma\!\left(\frac{u}{\tau};\frac{\sigma}{\tau},-\frac{1}{\tau}\right)}
	{\Gamma(u;\tau,\sigma)}.
	\label{eq:app-egamma-R}
\end{equation}
The Felder modular identity is equivalently
\begin{equation}
	R(u;\tau,\sigma)=e^{i\pi Q(u;\tau,\sigma)},
	\label{eq:app-egamma-RQ}
\end{equation}
where
\begin{align}
	Q(u;\tau,\sigma)
	&=
	\frac{u^3}{3\tau\sigma}
	-\frac{(\tau+\sigma-1)u^2}{2\tau\sigma}
	+\frac{\tau^2+\sigma^2+3\tau\sigma-3\tau-3\sigma+1}{6\tau\sigma}u
	\notag\\
	&\qquad
	+\frac{(\tau+\sigma-1)(\tau^{-1}+\sigma^{-1}-1)}{12}.
	\label{eq:app-egamma-Q}
\end{align}
Instead of taking this cubic phase as an input, we determine it from theta phases. Taking the quotient of~\eqref{eq:app-egamma-R} at adjacent Gamma lattice points and using~\eqref{eq:app-egamma-theta-shift-sigma}--\eqref{eq:app-egamma-theta-shift-tau} gives
\begin{align}
	\frac{R(u+\sigma;\tau,\sigma)}{R(u;\tau,\sigma)}
	&=
	\frac{\theta\!\left(\frac{u}{\tau};-\frac{1}{\tau}\right)}
	{\theta(u;\tau)}
	=
	e^{i\pi B(u,\tau)},
	\label{eq:app-egamma-Rdiff-sigma}\\
	\frac{R(u+\tau;\tau,\sigma)}{R(u;\tau,\sigma)}
	&=
	\frac{\theta\!\left(\frac{u}{\sigma};-\frac{1}{\sigma}\right)}
	{\theta(u;\sigma)}
	=
	e^{i\pi B(u,\sigma)}.
	\label{eq:app-egamma-Rdiff-tau}
\end{align}
Here $B$ is exactly the theta cocycle in~\eqref{eq:app-theta-S}. Therefore the Gamma phase obeys
\begin{align}
	Q(u+\sigma;\tau,\sigma)-Q(u;\tau,\sigma)&=B(u,\tau),
	\label{eq:app-egamma-Q-difference-sigma}\\
	Q(u+\tau;\tau,\sigma)-Q(u;\tau,\sigma)&=B(u,\sigma).
	\label{eq:app-egamma-Q-difference-tau}
\end{align}
This is the desired reduction: the finite differences of the Gamma phase are theta phases.

\subsubsection{Iteration and reconstruction}

Equations~\eqref{eq:app-egamma-Rdiff-sigma} and~\eqref{eq:app-egamma-Rdiff-tau} can be iterated. For any positive integer $N$,
\begin{align}
	\frac{R(u+N\sigma;\tau,\sigma)}{R(u;\tau,\sigma)}
	&=
	\prod_{j=0}^{N-1}
	\frac{\theta\!\left(\frac{u+j\sigma}{\tau};-\frac{1}{\tau}\right)}
	{\theta(u+j\sigma;\tau)},
	\label{eq:app-egamma-R-iterate-sigma}\\
	Q(u+N\sigma;\tau,\sigma)-Q(u;\tau,\sigma)
	&=
	\sum_{j=0}^{N-1}B(u+j\sigma,\tau).
	\label{eq:app-egamma-Q-iterate-sigma}
\end{align}
Similarly, iteration along the other elliptic period gives
\begin{align}
	\frac{R(u+N\tau;\tau,\sigma)}{R(u;\tau,\sigma)}
	&=
	\prod_{j=0}^{N-1}
	\frac{\theta\!\left(\frac{u+j\tau}{\sigma};-\frac{1}{\sigma}\right)}
	{\theta(u+j\tau;\sigma)},
	\label{eq:app-egamma-R-iterate-tau}\\
	Q(u+N\tau;\tau,\sigma)-Q(u;\tau,\sigma)
	&=
	\sum_{j=0}^{N-1}B(u+j\tau,\sigma).
	\label{eq:app-egamma-Q-iterate-tau}
\end{align}
Since $Q$ is a cubic polynomial in $u$, the two first-order finite-difference equations~\eqref{eq:app-egamma-Q-difference-sigma}--\eqref{eq:app-egamma-Q-difference-tau} determine its $u$-dependent coefficients from the theta cocycle $B$. The remaining additive constant is fixed by 
\begin{equation}
	Q\!\left(\frac{\tau+\sigma-1}{2};\tau,\sigma\right)=0.
\end{equation}
Thus the reconstruction of a Gamma modular phase is not a numerical fit and does not require learning a new phase rule. It can be collected by integration of theta phases.

\subsubsection{Product matching in the dataset}

The dataset contains products and quotients of Gamma factors, so one still has to identify where the ratio~\eqref{eq:app-egamma-R} appears. We write each factor schematically as $\Gamma(z;\tau,\sigma)^p$, with $p=+1$ for numerator factors and $p=-1$ for denominator factors. After cancelling identical numerator and denominator factors, the zero-phase moves used for matching are symmetry, periodicity, inversion, shifts, mod-1, and duplication:
\begin{align}
	\Gamma(z;\tau,\sigma)&=\Gamma(z;\sigma,\tau),\\
	\Gamma(z+n;\tau+m,\sigma+\ell)&=\Gamma(z;\tau,\sigma),\qquad n,m,\ell\in \mathbb{Z},\\
	\Gamma(z;\tau,\sigma)&=\frac{1}{\Gamma(\tau+\sigma-z;\tau,\sigma)},\\
	\Gamma(z;\tau,\sigma)&=\frac{1}{\Gamma(z-\tau;-\tau,\sigma)},\\
	\Gamma(z;\tau,\sigma)&=\Gamma(\sigma-z;-\tau,\sigma),\\
	\Gamma(z;\tau,\sigma)&=\Gamma(z;\tau-\sigma,\sigma)\Gamma(z;\sigma-\tau,\tau),\\
	\Gamma(z;\tau,\sigma)&=
	\prod_{a,b,c\in\{0,1\}}
	\Gamma\!\left(\frac{z+a\tau+b\sigma+c}{2};\tau,\sigma\right).
\end{align}
These moves only identify and align factors. The only phase-bearing pattern is the Felder ratio
\begin{align}
	&\Gamma\!\left(\frac{z}{\sigma};\frac{\tau}{\sigma},-\frac{1}{\sigma}\right)^p
	\Gamma\!\left(\frac{z}{\tau};\frac{\sigma}{\tau},-\frac{1}{\tau}\right)^p
	\Gamma(z;\tau,\sigma)^{-p}
	\notag\\
	&\hspace{4cm}
	=
	e^{i\pi p Q(z;\tau,\sigma)}.
	\label{eq:app-egamma-ratio-pattern}
\end{align}
Once this pattern is identified, its phase is computed from the theta finite differences above.

\subsubsection{Worked example corresponding to Table~\ref{tab:egam}}

We now apply this ratio method to the example in Table~\ref{tab:egam}. The complicated expression contains the following three factors:
\begin{align}
&\Gamma\!\left(-\frac{2z}{4\tau+7};
	\frac{-\sigma-28\tau-49}{4\tau+7},
	\frac{3\tau+5}{4\tau+7}\right)
\Gamma\!\left(\frac{2z}{\sigma+28\tau+49};
	\frac{-4\tau-7}{\sigma+28\tau+49},
	\frac{-3\tau-5}{\sigma+28\tau+49}\right)
\notag\\
&\qquad\times
\Gamma\!\left(
	\frac{2z}{3\tau+5};
	\frac{\sigma+28\tau+49}{3\tau+5},
	\frac{-4\tau-7}{3\tau+5}
	\right)^{-1}.
	\label{eq:app-egamma-example-ratio}
\end{align}
Set
\begin{equation}
	u=\frac{2z}{3\tau+5},\qquad
	\alpha=\frac{\sigma+28\tau+49}{3\tau+5},\qquad
	\beta=\frac{-4\tau-7}{3\tau+5}.
\end{equation}
Then~\eqref{eq:app-egamma-example-ratio} is precisely the ratio $R(u;\alpha,\beta)$. Therefore its phase is determined by the theta ratios
\begin{align}
	\frac{R(u+\beta;\alpha,\beta)}{R(u;\alpha,\beta)}
	&=
	\frac{\theta\!\left(\frac{u}{\alpha};-\frac{1}{\alpha}\right)}
	{\theta(u;\alpha)}
	=
	e^{i\pi B(u,\alpha)},
	\label{eq:app-egamma-example-theta-alpha}\\
	\frac{R(u+\alpha;\alpha,\beta)}{R(u;\alpha,\beta)}
	&=
	\frac{\theta\!\left(\frac{u}{\beta};-\frac{1}{\beta}\right)}
	{\theta(u;\beta)}
	=
	e^{i\pi B(u,\beta)}.
	\label{eq:app-egamma-example-theta-beta}
\end{align}
Equivalently, the exponent $P(z,\tau,\sigma)=Q(u;\alpha,\beta)$ is recovered from
\begin{align}
	Q(u+\beta;\alpha,\beta)-Q(u;\alpha,\beta)&=B(u,\alpha),\\
	Q(u+\alpha;\alpha,\beta)-Q(u;\alpha,\beta)&=B(u,\beta),
\end{align}
where the right-hand sides are computed by the theta reconstruction of Appendix~\ref{app:qtheta-phase}. The remaining factors in Table~\ref{tab:egam} are related to the simplified expression by the zero-phase moves listed above: the eight denominator factors form a duplication orbit, one remaining factor is aligned by symmetry, and another is aligned by period shifts. These steps identify the ratio~\eqref{eq:app-egamma-example-ratio}; they do not contribute additional phase.

Integrating the two theta differences and using $Q((\alpha+\beta-1)/2;\alpha,\beta)=0$ gives
\begin{align}
	P(z,\tau,\sigma)
	&=
	\frac{1}{
	144\sigma\tau^2+492\sigma\tau+420\sigma
	+4032\tau^3+20832\tau^2+35868\tau+20580
	}
	\notag\\
	&\quad\times
	\Bigl(
	-7\sigma^2\tau-4\sigma^2z-12\sigma^2
	-331\sigma\tau^2-140\sigma\tau z-1149\sigma\tau
	\notag\\
	&\qquad
	+24\sigma z^2-248\sigma z-997\sigma
	-3864\tau^3-1028\tau^2z-20206\tau^2
	\notag\\
	&\qquad
	+504\tau z^2-3664\tau z-35219\tau
	-32z^3+888z^2-3264z-20461
	\Bigr).
	\label{eq:app-egamma-phase-example}
\end{align}
Thus the two expressions in Table~\ref{tab:egam} satisfy
\begin{equation}
	F_{\mathrm{complicated}}=e^{i\pi P(z,\tau,\sigma)}F_{\mathrm{simple}}.
\end{equation}

\bibliographystyle{JHEP}
\bibliography{bib-bh}
 
\end{document}